\documentclass{IEEEtran}

\usepackage[colorlinks=true,
            urlcolor=blue,
            linkcolor=blue,
            citecolor=blue]{hyperref}

\usepackage{cite}
\usepackage{amsmath,amssymb,amsfonts}
\usepackage{graphicx}
\usepackage{textcomp}
\usepackage{color}
\usepackage[normalem]{ulem}
\usepackage{float}

\begin{document}

\title{Interference-Tolerant Mixer-First Receivers for FR3: Design Principles and Tradeoffs}

\author{
Reza Nikandish~\IEEEmembership{Senior Member,~IEEE} and
Hamed Rahmani~\IEEEmembership{Senior Member,~IEEE}
\thanks{The authors are with the Department of Electrical and Computer Engineering,
New York University, New York, NY, USA.}
\thanks{This work was supported by the National Telecommunications and Information Administration (NTIA) under Grant 3660IF2419.}
}

\maketitle

\begin{abstract}
Frequency Range 3 (FR3), spanning 7.125--24.25~GHz, is a promising candidate band for 6G communications, bridging the coverage of sub-6~GHz bands and the capacity of millimeter-wave frequencies. Its incumbent-dense and non-contiguous spectrum demands frequency-agile, interference-tolerant receivers (RXs) capable of hopping across fragmented sub-bands while withstanding strong blockers. Mixer-first RXs are well suited to this role, yet their design for FR3 has not been systematically addressed. This paper presents a hardware design perspective on mixer-first RXs for FR3, evaluating selectivity enhancement, harmonic rejection, linearization, low-noise design, and multi-phase clock generation under FR3-specific constraints. The analysis identifies viable techniques, fundamental limitations, and circuit- and architecture-level design tradeoffs. A central insight is that the frequency-translational property of mixer-first RXs allows selectivity, linearization, and noise cancellation to be implemented partly at baseband and translated to RF, enabling frequency-agile FR3 operation while shifting the dominant design constraints to mixer parasitics, baseband circuit robustness, and multi-phase clock generation.
\end{abstract}

\begin{IEEEkeywords}
6G, blocker rejection, frequency range 3 (FR3), harmonic rejection, mixer-first receiver, N-path filter, noise figure, receiver linearity.
\end{IEEEkeywords}

\section{Introduction}
\label{section:intro}

\begin{figure}[!t]
  \centering
  \includegraphics[width=\columnwidth]{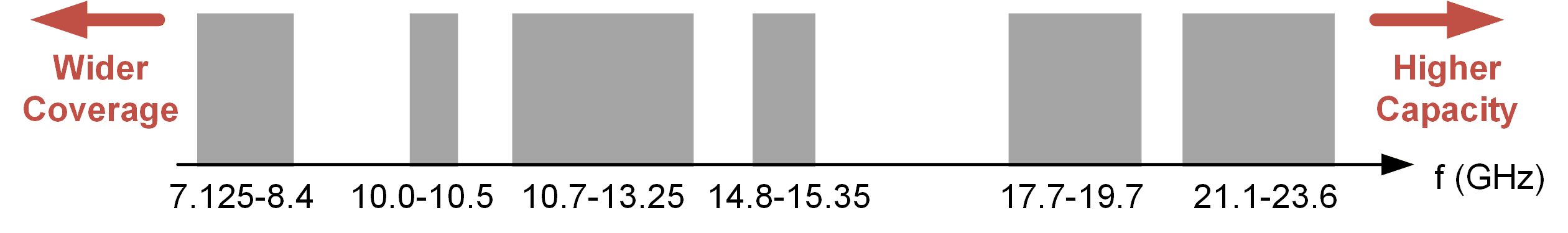}
  \caption{FR3 providing a balance between the wide coverage of FR1 and the high capacity of FR2, and its fragmented candidate sub-bands.}
  \label{fig:FR3}
\end{figure}

Frequency Range 3 (FR3), spanning 7.125--24.25~GHz, is being explored for future wireless systems because it offers a compromise between the coverage available below 7~GHz and the bandwidth available at millimeter-wave frequencies \cite{Rappaport_npjwt_2025, Rappaport_access_2013, razavi_sscs_2023}. The U.S. National Telecommunications and Information Administration has identified 7.125--8.4~GHz for initial studies \cite{NTIA_2024}, although 6G allocations have not been finalized. Much of FR3 is occupied by satellite, radar, radio-astronomy, and other incumbent services, leaving candidate cellular spectrum fragmented across non-contiguous sub-bands, as illustrated in Fig.~\ref{fig:FR3}. Receivers (RXs) must therefore combine a wide tuning range with strong interference tolerance. The same frequency agility can also support multiband aggregation for integrated sensing and communication \cite{Pegoraro_arxiv_2025}.

Mixer-first RXs provide a frequency-agile alternative to fixed or narrowly tunable RF filters. Their center frequency is controlled by the local oscillator (LO), while the switching network translates a baseband (BB) impedance to the RF input \cite{Andrews_jssc_2010, Andrews_tcas1_2010, Klumperink_cicc_2017}. This property can provide input matching and channel selectivity without an RF low-noise amplifier (LNA) preceding the mixer. It also exposes the antenna to switch loss and parasitic capacitance and places stronger dynamic-range requirements on the BB circuitry. Multiphase LO generation and LO-to-antenna leakage introduce additional constraints. Therefore, mixer-first RX is a very promising architecture for FR3, but its suitability must be evaluated from these coupled RF, BB, and clock (CK) tradeoffs.

Prior work has addressed the main limitations of mixer-first RXs through circuit- and architecture-level techniques. Higher-order BB impedance synthesis has been used to translate sharper BB filtering responses to RF, while cascaded and gain-boosted N-path filters realize higher-order filtering directly through multiple frequency-translational stages \cite{Nauta_isscc_2025, yang_MWM_2026, Klumperink_cicc_2017, Jain_jmw_2021, Lien_jssc_2018, Krishnamurthy_jssc_2020, Krishnamurthy_sscl_2020, Krishnamurthy_sscl_2021, Pini_jssc_2020, Zanten_jssc_2025, darvishi_jssc_2013, Song_jssc_2021, razavi_jssc_2022, Ye_jssc_2025}. Harmonic blockers have been addressed using BB harmonic recombination, weighted RF paths, switched-capacitor harmonic-rejection mixers, and harmonic traps \cite{Andrews_jssc_2010, araei_jssc_2023, Weinreich_jssc_2023, araei_jssc_2024, razavi_jssc_2022}. Mixer linearity has been improved through bottom-plate mixing and switch clock boosting or bootstrapping, while the BB amplifier can be linearized through feedback and virtual-ground operation \cite{Lien_jssc_2019, araei_jssc_2026, Hardeveld_jssc_2026, Krishnamurthy_jssc_2021}. The NF penalty associated with removing RF gain before the mixer has motivated noise-canceling and gain-assisted architectures \cite{Murphy_jssc_2012, Murphy_jssc_2015, wu_jssc_2015, Bhat_jssc_2021, Kumar_jssc_2026, Wu_tmtt_2016, Wu_jssc_2025}.

Most of these techniques, however, have been demonstrated below FR3, and their performance cannot be extrapolated based on operating frequency alone. In particular, the mixer switch introduces a fundamental tradeoff between on-resistance and parasitic capacitance: increasing the switch width reduces $R_\mathrm{sw}$ and its associated loss and noise, but increases the capacitance at the RF input and the capacitive load presented to the LO driver. These effects become more significant as frequency increases. In addition, the wider channel bandwidths considered for FR3 increase the gain-bandwidth, noise, linearity, and dynamic range requirements of the active BB circuits. Therefore, extending mixer-first RXs to FR3 requires considering the interactions among RF loss and parasitics, BB circuit requirements, blocker tolerance, and LO generation rather than evaluating the individual circuit techniques independently.

This paper evaluates mixer-first RX techniques against
FR3-specific constraints, including wide fractional bandwidth incumbent-dense spectrum, wider modulation bandwidth, and stronger parasitic sensitivity. For the key design challenges of selectivity enhancement, harmonic rejection, linearization, low-noise design, and multi-phase clock generation, the paper
identifies which techniques remain viable in FR3, which face
fundamental limitations, and which require new circuit or
architecture-level solutions. A central insight is to exploit the
frequency translational property of the N-path mixer-first RX
as well as the transparency of mixer switches, to enhance
the channel selectivity, linearity, and blocker tolerance of the
RX chain. The resulting analysis establishes design guidelines
for frequency-agile, interference-tolerant RXs across the FR3
bands and highlights the open problems that remain for future
research.

The paper is organized as follows. Section~\ref{section:fundamentals} establishes the interference metrics and mixer-first operating principles. Sections~\ref{section:selectivity}--\ref{section:noise} address selectivity, harmonic rejection, linearity, and noise. Section~\ref{section:LO} discusses multiphase LO generation and leakage, and Section~\ref{section:FR3_insights} summarizes the design guidelines and open problems.

\section{Fundamentals of Blocker-Tolerant Receivers}
\label{section:fundamentals}

\subsection{Blockers}

In wireless communications, the RX must reliably detect a weak desired signal in the presence of strong interfering signals, collectively referred to as blockers. In FR3, this challenge is particularly severe because the segmented spectrum places incumbent satellite, radar, and other services in close proximity to the operating channel, producing blockers spanning a wide range of frequencies and power levels. As illustrated in Fig.~\ref{fig:spectrum_fundamentals}, blockers affecting the RX can be classified into three categories based on their frequency relationship to the desired channel: in-band (IB), out-of-band (OOB), and harmonic blockers.

\subsubsection{In-Band Blockers} 
IB blockers reside within the channel band, as shown in Fig.~\ref{fig:spectrum_fundamentals}(a), and are down-converted to BB through reciprocal mixing with noisy LO sidebands. These blockers cannot be directly filtered, thus imposing stringent requirements on RX linearity and CK jitter. This challenge becomes even more severe for the wide modulation bandwidths, e.g., 400 MHz, expected in FR3.

\subsubsection{Out-of-Band Blockers} 
OOB blockers can be close-in or far-out relative to the channel, as shown in Fig. \ref{fig:spectrum_fundamentals}(a). A conventional mixer-first RX exhibits a first-order 20 dB/dec roll-off that is often insufficient to suppress close-in OOB blockers. Therefore, as shown in Fig. \ref{fig:spectrum_fundamentals}(b), selectivity must be enhanced, which is discussed in Section \ref{section:selectivity}.

\subsubsection{Harmonic Blockers} 
\begin{figure}[!t]
  \centering
  \includegraphics[width=0.7\columnwidth]{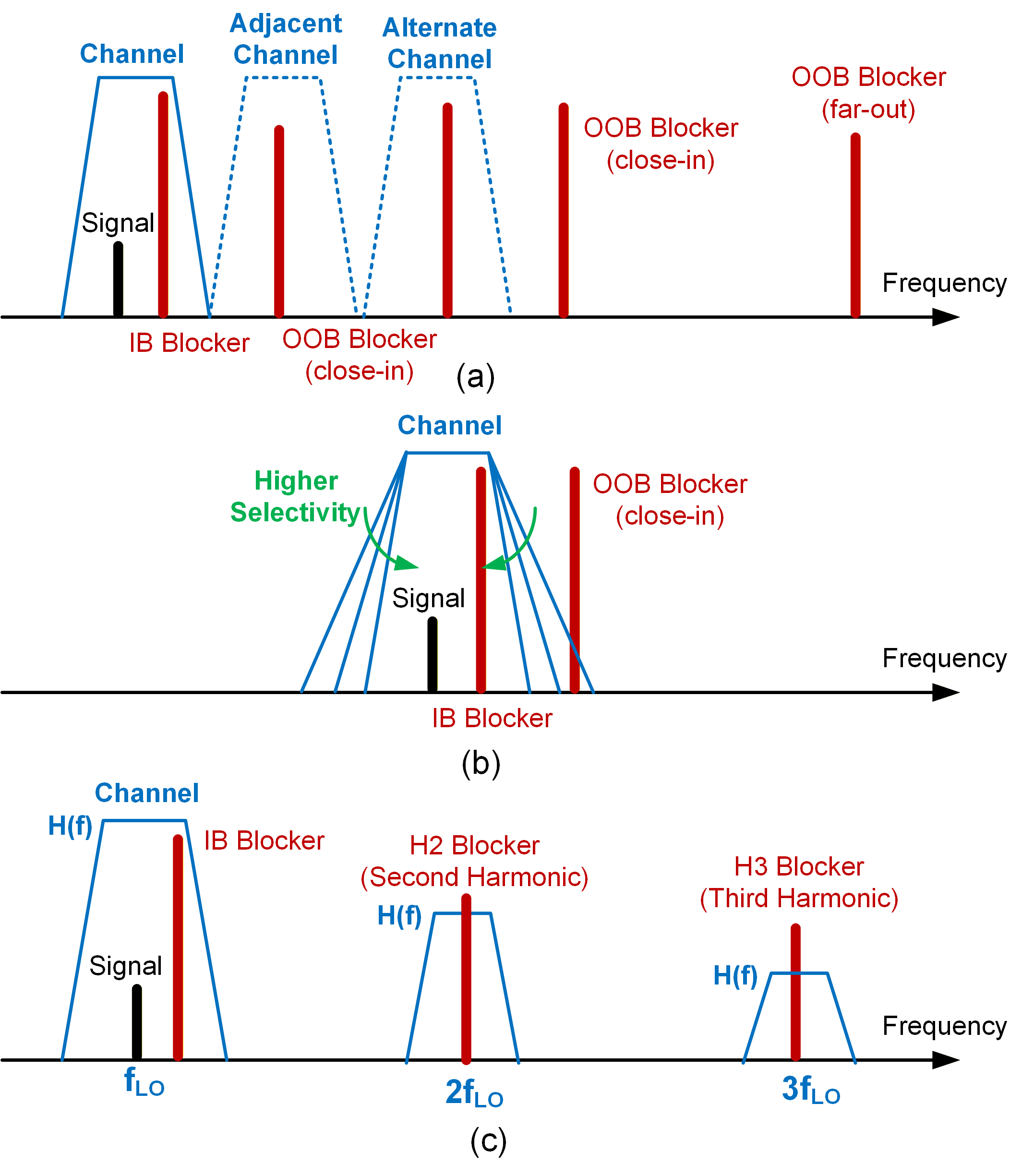}
  \caption{The concepts of (a) in-band (IB) and out-of-band 
  (OOB) blockers, close-in and far-out OOB blockers, 
  (b) selectivity, (c) harmonic blockers.}
  \label{fig:spectrum_fundamentals}
\end{figure}

Harmonic blockers arise from the harmonic content of the square-wave LO, which causes signals at $nf_{\rm LO}$ to be down-converted to BB alongside the desired signal, as shown in Fig. \ref{fig:spectrum_fundamentals}(c). Harmonic rejection techniques are discussed in Section \ref{section:harmonic}.

\subsection{Mixer-First Receivers}
\begin{figure}[!t]
  \centering
  \includegraphics[width=0.5\columnwidth]{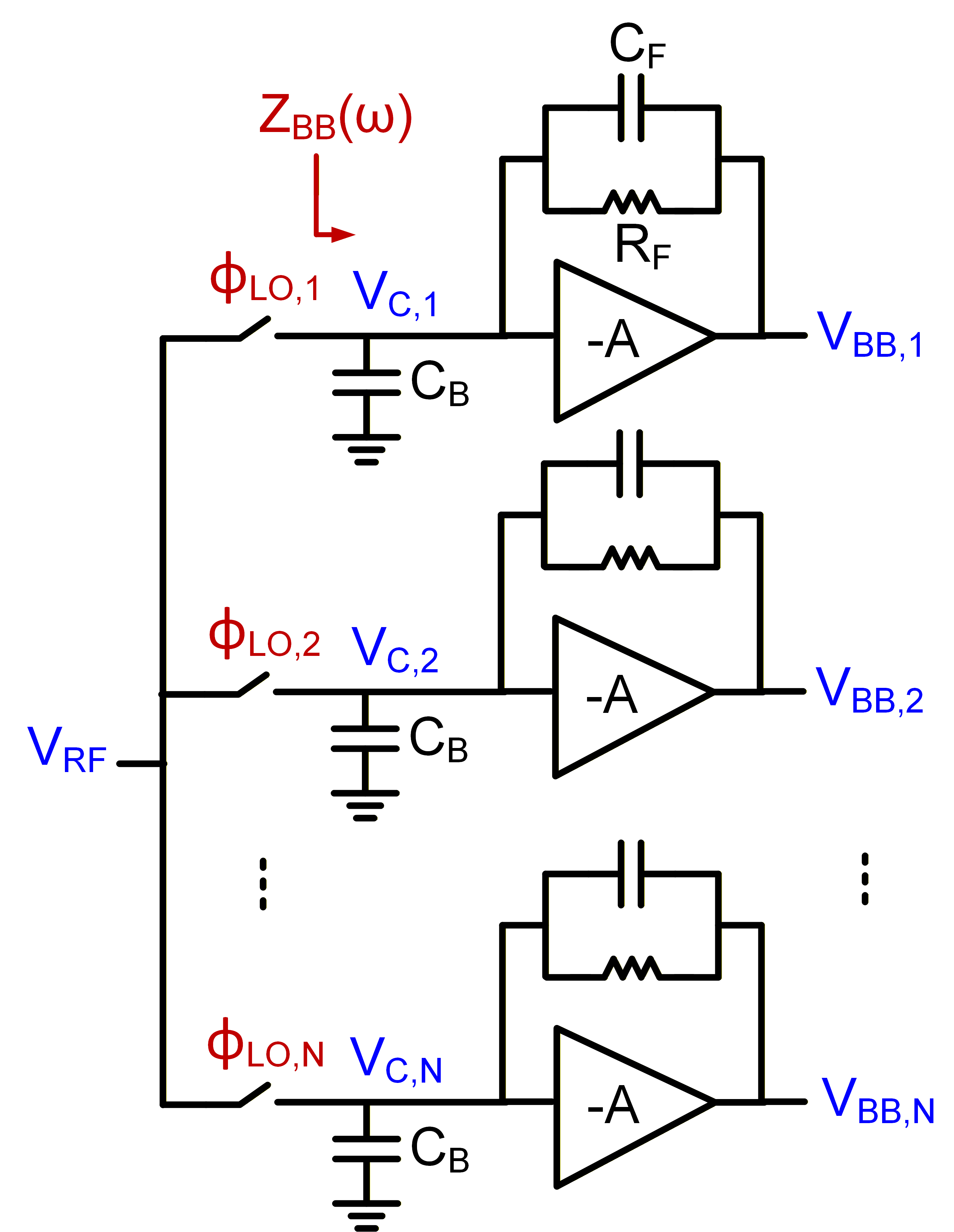}
  \caption{Mixer-first receiver architecture using 
  non-overlapping LO phases $\phi_{\rm LO,\{1...N\}}$. 
  The impedance $Z_{BB}(\omega)$ refers to the impedance 
  presented to BB ports of the mixers.}
  \label{fig:MFRX}
\end{figure}

The mixer-first RX is particularly well suited to FR3 among RX architectures because its frequency translational operation enables RF selectivity, harmonic rejection, and noise cancellation to be realized at BB frequencies.  alleviating the parasitic, tuning, and scalability limitations of direct RF-domain implementation. In contrast, architectures relying on active RF gain stages, e.g., LNA-first RX~\cite{Borremans_jssc_2011, xu_jssc_2018}, low-noise transconductance amplifier (LNTA)-first RX~\cite{ru_jssc_2009}, and gain-boosted N-path RX~\cite{park_jssc_2014, razavi_jssc_2022, lin_tcas1_2014, lin_jssc_2014, araei_jssc_2026}, become increasingly sensitive to parasitic capacitance at higher frequencies and struggle to provide sufficient blocker tolerance across the FR3 bands.

The mixer-first RX architecture is shown in Fig.~\ref{fig:MFRX}, 
where N-path mixers driven by N-phase non-overlapping LO phases 
realize RF channel filtering. The BB amplifiers amplify the 
out-of-phase voltages across the sampling capacitors, $V_{C,k}$, 
using RC feedback for low-pass filtering. The number of 
paths $N$ trades off noise performance, harmonic rejection, and 
CK generation complexity. $N=4$ yields a simpler four-phase 
25\% duty-cycle CK, while $N=8$ can achieve lower NF 
(Section~\ref{section:noise}) and enables harmonic rejection 
(Section~\ref{section:harmonic}) at the cost of a more complex 
eight-phase 12.5\% duty-cycle CK operating at a higher input 
frequency (Section~\ref{section:LO}). The phase count must therefore be selected jointly with the NF, harmonic-rejection, conversion-loss, and clock-power targets. Four paths reduce clock complexity, whereas eight paths directly support several harmonic-recombination schemes; neither choice is universally optimum in FR3.

The operation of the mixer-first RX is fundamentally controlled 
by the BB impedance $Z_{BB}(\omega)$ through the impedance 
transparency property of the mixers: the impedance seen at the 
RF port is a frequency-translated version of $Z_{BB}$, given by 
$Z_{BB}(\omega_{LO}-\omega)$ ~\cite{Andrews_jssc_2010, 
Andrews_tcas1_2010, yang_tcas1_2015}. The input impedance of the mixer-first RX circuit in Fig. \ref{fig:MFRX} can 
be expressed as
\begin{equation}
    \label{Zin}
    Z_{in}(\omega) =
\frac{1}{j\omega C_p}
\parallel
\left[
R_{sw}
+
Z_{sh}(\omega)
\parallel
\gamma_N Z_{BB}(\omega_{LO}-\omega)
\right]
\end{equation}
\begin{equation}
    \label{gamma}
    \gamma_N = \frac{1}{N}\,{\rm sinc}^2\!\left(\frac{\pi}{N}
    \right),
\end{equation}
where $C_p$ is the switch parasitic capacitance, $R_{sw}$ is 
the switch on-resistance, and $Z_{sh}(\omega)$ models harmonic 
losses~\cite{Andrews_tcas1_2010, yang_tcas1_2015, 
Mirzaei_tcas1_2010}. At low frequencies where $C_p$ and 
$Z_{sh}$ can be neglected, the input matching condition around $f_{RF} \approx f_{LO}$ can be simplified to 
\begin{equation}
    \label{RS}
    R_S \approx R_{sw} + \gamma_N Z_{BB}(0).
\end{equation}
This condition is usually satisfied by setting the low-frequency component of the BB impedance, $Z_{BB}(0) = R_{BB}$. At higher frequencies of FR3, however, the larger $C_p$ associated with 
wider switches degrades the input reflection coefficient $S_{11}$ and introduces additional 
loss. The importance of this loading is set by the dimensionless product $\omega R_S C_p$, rather than by frequency alone.
When $f_{RF}$ is far from $f_{LO}$, the input impedance is determined by $Z_{BB}(\omega_{LO} - \omega_{RF})$ which has a small value, resulting in $Z_{in}(\omega) \approx R_{sw}$, which is usually much smaller than the source resistance $R_S$. Therefore, outside the channel bandwidth, 
$S_{11} \approx 0$~dB, reflecting OOB blockers and 
suppressing their transfer into the RX.

The BB impedance controls the channel bandwidth such that RF signals within the channel bandwidth can pass through the RF front-end, while blockers at other frequencies are attenuated. The channel bandwidth (BW) can be defined as
\begin{equation}
    \label{BW}
    \left|f_{RF} - f_{LO} \right| < \frac{1}{2}BW,
\end{equation}
which is twice the BB bandwidth, $ BW = 2f_{BB}$, arising from the BB to RF translation. The channel bandwidth is primarily determined by the capacitance $C_B$ and $C_F$ in Fig.~\ref{fig:MFRX}. 

\subsection{Switch Sizing and Frequency Scaling}

To first order, the on-resistance and parasitic capacitances of a MOS switch scale with its width $W$ as
\begin{equation}
 R_{sw}(W)=  R_0\frac{W_0}{W},\qquad
 C_p(W)=  C_{0}\frac{W}{W_0},
 \label{eq:sw_scaling}
\end{equation}
where $R_0$ and $C_{0}$ denote values at a reference width $W_0$. Increasing $W$ reduces conduction loss and the thermal noise contribution of the switch, but increases loss due to the parasitic capacitance. Furthermore, the switches gate capacitance $C_g$ increases with their width, which requires higher power consumption in the LO driver circuits. A first-order estimate of the LO driver power can be expressed as
\begin{equation}
 P_{LO} \approx \alpha N C_{g}(W) f_{CK} V_{CK}^{2},
 \label{eq:lo_power}
\end{equation}
where $\alpha$ includes switching activity and driver overhead. Equations~(\ref{Zin}), (\ref{eq:sw_scaling}), and (\ref{eq:lo_power}) expose a three-way tradeoff: a wider switch lowers $R_{sw}$ but increases both $\omega R_SC_p$ and LO power. An optimum width can therefore exist for a specified frequency, matching requirement, NF target, and clock-power budget. Its value is technology- and architecture-dependent; the common rule of minimizing $R_{sw}$ alone is insufficient in FR3.

\subsection{From Selectivity to Blocker Tolerance}

Input mismatch and channel filtering reduce the blocker power delivered to internal nodes, but they do not alone determine blocker tolerance. For a blocker at offset $\Delta f$, let $H_k(\Delta f)$ denote the transfer from the antenna to node $k$. The maximum input blocker consistent with linear operation is bounded by
\begin{equation}
 P_{B,max}(\Delta f)=\min_k\left\{P_{k,lim}-20\log_{10}|H_k(\Delta f)|\right\},
 \label{eq:blocker_limit}
\end{equation}
when powers and gains are expressed in dB units. $P_{k,lim}$ denotes the upper limit of permitted power level at node $k$. Relevant limits include switch compression, sampling-node voltage, BB feedback amplifier linearity, and BB output voltage swing. Higher-order selectivity improves $H_k$ for OOB blockers, but the improvement in input-referred blocker tolerance stops when another node or reciprocal mixing becomes limiting factor. Meaningful comparisons should therefore specify blocker power, normalized offset $\Delta f/BW$, desired signal level, gain state, and the metric being measured, such as B1dB, IM3, cross-modulation, or blocker NF.

\subsection{LO Leakage to Antenna}

In a mixer-first RX, elimination of the LNA also removes the reverse isolation that normally separates the LO-driven mixer from the antenna. Clock feed-through by switch overlap capacitance can produce an LO component at the RF port. This leakage can couple between antennas or channels and can interference with neighboring receivers. A 28-nm CMOS mixer-first RX demonstrated wideband leakage suppression using an embedded multi-bit mixer digital-to-analog converter (DAC) to enhance device matching \cite{Wu_tmtt_2016_LO}. The the LO-to-antenna leakage can be suppressed using differential switching, symmetric layout, balanced transformers, filtering, and digitally controlled mixer-device tuning. The digital approaches provide high flexibility and can be used in FR3 bands. 


\section{Selectivity Enhancement}
\label{section:selectivity}

In a conventional mixer-first RX, the BB impedance is typically approximated as a parallel RC 
network. In the circuit of Fig.~\ref{fig:MFRX}, assuming that 
the BB amplifier provides a constant gain within 
the channel bandwidth, $Z_{BB}(\omega)$ consists of the 
resistance $R_{BB}=R_F/(1+A)$ in parallel with the capacitance 
$C_{BB}=C_B+(1+A)C_F$. This BB impedance, through frequency translation, produces a first-order RF transfer function around the LO frequency with a 20~dB/dec roll-off. Such selectivity is often insufficient to suppress close-in OOB blockers. Therefore, sharper RF selectivity must be synthesized through either higher-order BB impedance or higher-order N-path filters.

\subsection{Higher-Order BB Impedance}

\begin{figure}[!t]
  \centering
  \includegraphics[width = 0.6\columnwidth]{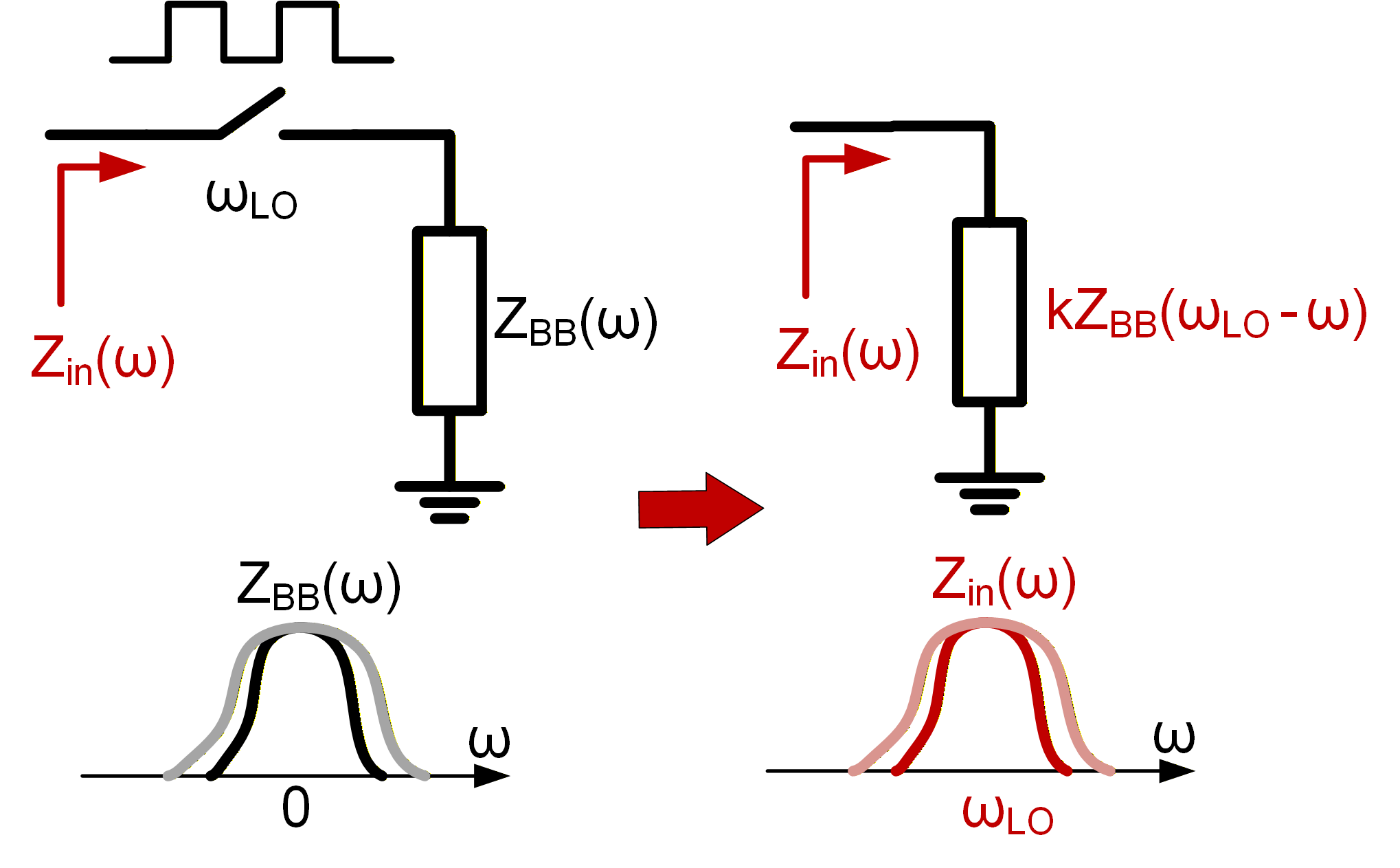}
  \caption{Translation of the BB impedance to the LO frequency by 
  the mixer, indicating that a higher-order BB impedance can 
  provide sharper RF roll-off and enhance the selectivity of the 
  mixer-first RX.}
  \label{fig:Higher_Order_ZBB}
\end{figure}

A direct approach to improving selectivity is to synthesize a higher-order BB impedance. As shown in Fig.~\ref{fig:Higher_Order_ZBB}, the mixer translates this response to the LO frequency. The active synthesis circuit operates at BB and therefore avoids additional RF resonators or cascaded RF switches. The RF input still contains the mixer resistance and capacitance in (\ref{Zin}), while the BB circuit must meet noise, stability, gain-bandwidth, and blocker-current requirements. The impedance must also satisfy the input-matching condition.

\begin{figure}[!t]
  \centering
  \includegraphics[width = 0.8\columnwidth]{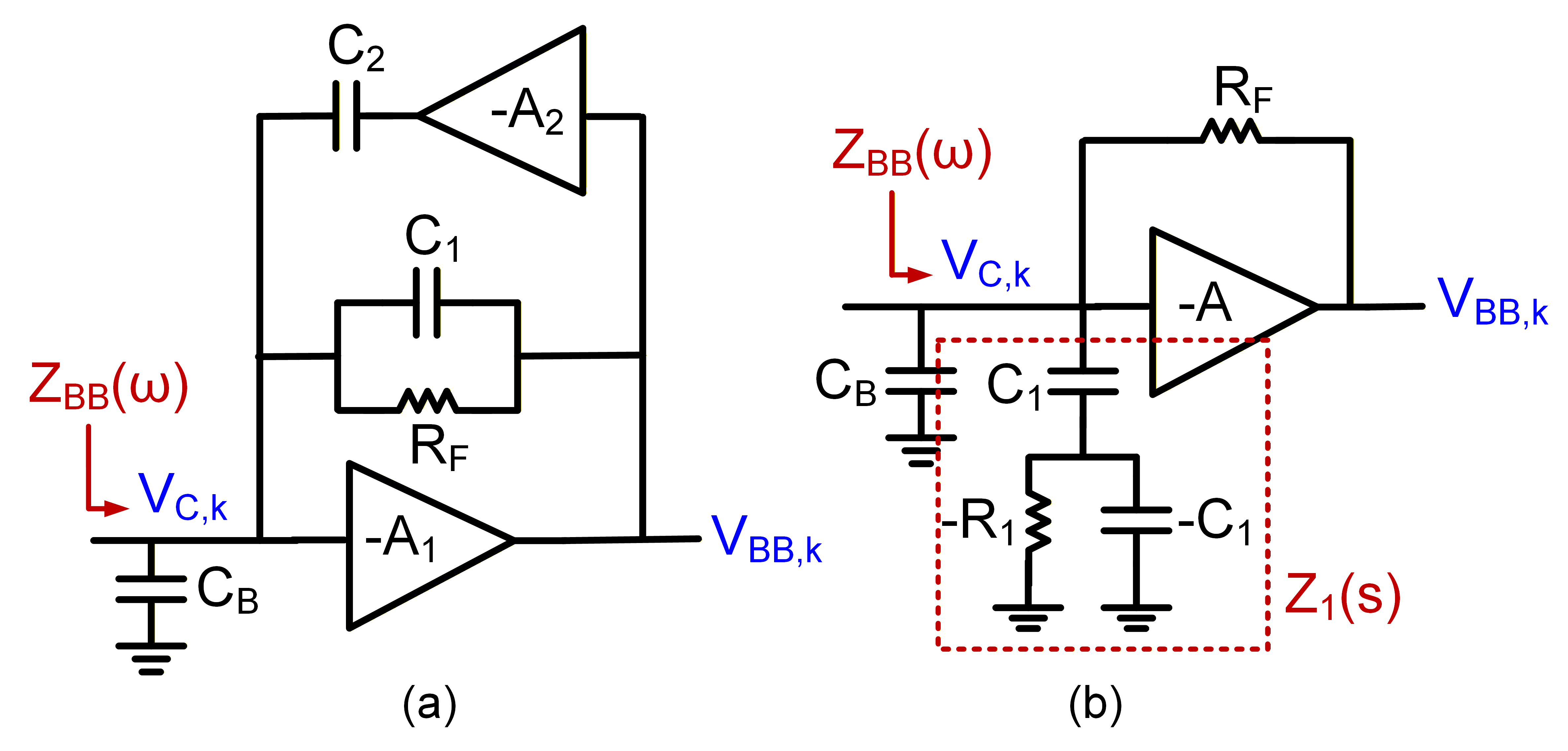}
  \caption{(a) Higher-order BB impedance realized using a dual 
  negative/positive feedback network~\cite{Lien_jssc_2018}. 
  (b) Higher-order dual-pole BB impedance realized using an 
  active negative RC network~\cite{Krishnamurthy_jssc_2020}.}
  \label{fig:Higher_Order_ZBB_CKT}
\end{figure}

A circuit for synthesizing a higher-order BB impedance is shown 
in Fig.~\ref{fig:Higher_Order_ZBB_CKT}(a), where dual negative 
and positive feedback networks realize an impedance with 
sharper roll-off~\cite{Lien_jssc_2018}. The resulting BB 
admittance can be expressed as
\begin{equation}
    \label{ZBB_1}
    Z_{BB}^{-1}(s) = C_B s + Y_N(s) + Y_P(s),
\end{equation}
where the negative feedback path produces the conventional RC 
response
\begin{equation}
    \label{YBBN}
    Y_N(s) \approx (1+A_1)C_1 s + \frac{1+A_1}{R_F},
\end{equation}
whereas the positive feedback path introduces an additional 
frequency-dependent term
\begin{equation}
    \label{YBBP}
    Y_P(s) \approx C_1 C_2 R_o s^2 + (1-A_1A_2)C_2 s.
\end{equation}
The $s^2$ term effectively realizes a super-capacitor 
response that sharpens the selectivity roll-off. This approach 
has demonstrated 40~dB/dec selectivity enhancement together 
with wideband frequency tuning across 0.2--8.0 GHz \cite{Lien_jssc_2018}.

Another approach for realizing a higher-order BB impedance is 
shown in Fig.~\ref{fig:Higher_Order_ZBB_CKT}(b), where a 
dual-pole impedance is synthesized using a negative RC network 
\cite{Krishnamurthy_jssc_2020}. While an all-pole impedance is 
not realizable using only passive elements, the negative 
elements $-R_1$ and $-C_1$ enable the realization of the 
dual-pole response
\begin{equation}
    \label{Z1}
    Z_1(s) = \frac{1}{sC_1(1+sR_1C_1)}.
\end{equation}
The resulting BB admittance can be expressed as
\begin{equation}
    \label{ZBB_2}
    Z_{BB}^{-1}(s) = R_1C_1^2 s^2 + (C_1 + C_B)s + \frac{1+A}{R_F},
\end{equation}
where the $s^2$ term enhances the roll-off of the BB impedance and, as a result, the translated RF selectivity. Experimental implementations of 
higher-order BB impedance techniques have demonstrated 
selectivity roll-offs ranging from 40~dB/dec to 80~dB/dec, 
together with wideband operation and strong close-in blocker 
rejection~\cite{Krishnamurthy_jssc_2020, 
Krishnamurthy_sscl_2020, Krishnamurthy_sscl_2021, 
Pini_jssc_2020}.

A common challenge in the circuits of 
Fig.~\ref{fig:Higher_Order_ZBB_CKT} is maintaining stability 
in the presence of multiple feedback loops. Prior work has 
addressed this issue using additional stabilization networks 
\cite{Lien_jssc_2018, Krishnamurthy_jssc_2020}, though the 
problem becomes more severe as higher-order responses are used 
to achieve sharper selectivity. Another important limitation is the noise contribution of the active feedback circuits, which can degrade the NF of the RX. In addition, strong blockers can drive the active circuits into nonlinear operation, causing the BB impedance to deviate from its intended response and reducing blocker rejection.

Higher-order BB impedance synthesis reduces the number of frequency-selective elements operating directly at RF. This advantage is most useful when the added BB feedback network can realize the required channel bandwidth without excessive noise or power and remains linear under the residual blocker current. Its FR3 suitability must therefore be established jointly with the mixer parasitics and BB dynamic range, not inferred from the BB operating frequency alone.

\subsection{Higher-Order N-Path Filter}

\begin{figure}[!t]
  \centering
  \includegraphics[width =0.6\columnwidth]{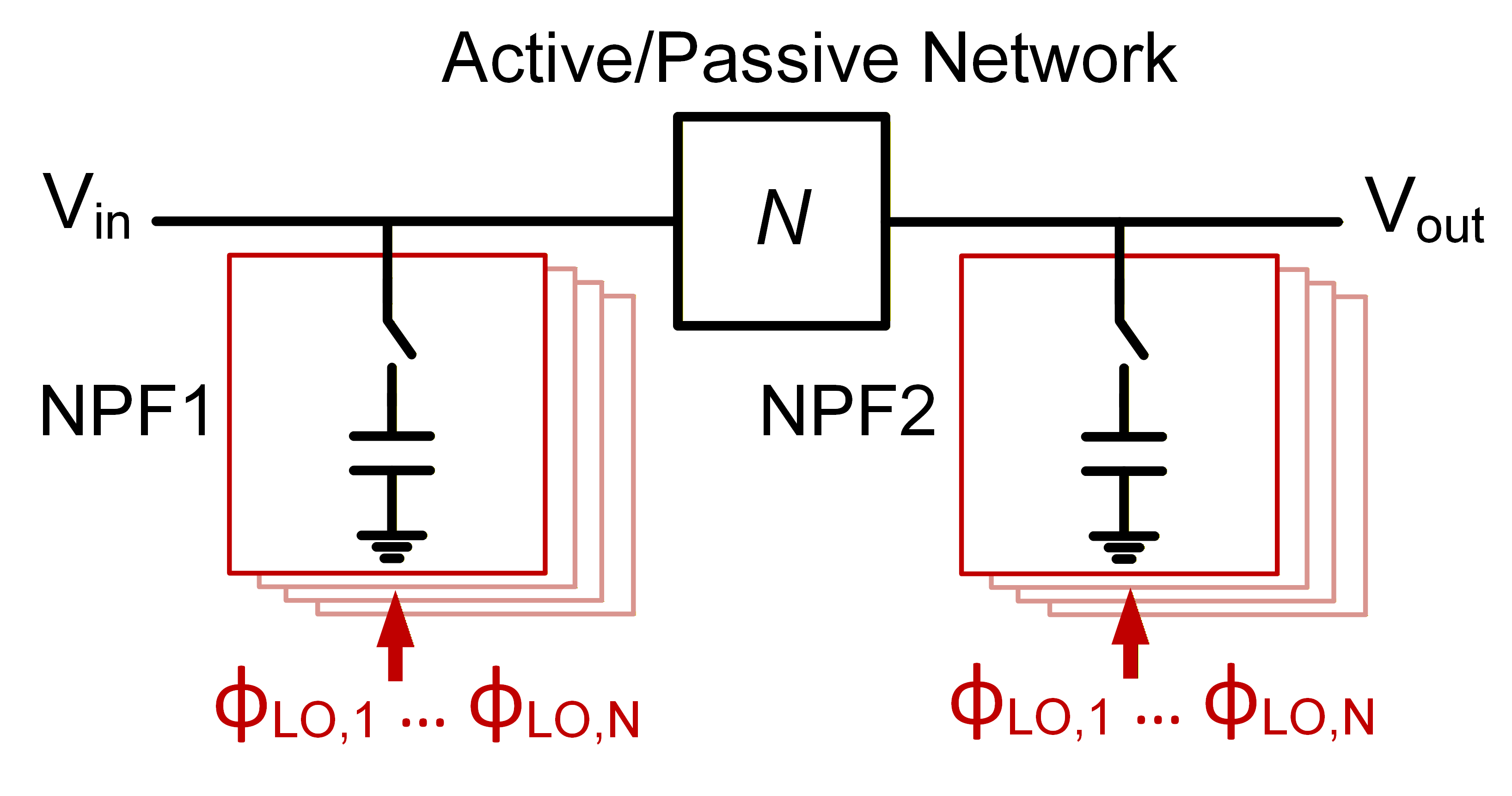}
  \caption{Higher-order N-path filter with an isolation network 
  between two N-path filter sections. The isolation network can 
  be realized using either active or passive circuits.}
  \label{fig:Higher_Order_NPF}
\end{figure}

Another approach for enhancing selectivity is to realize a 
higher-order RF transfer function by cascading multiple N-path 
filter sections. However, unlike conventional linear 
time-invariant (LTI) filters, N-path filters are linear time-varying 
(LTV) circuits whose operation can be significantly altered by 
direct loading between adjacent stages. Therefore, isolation 
networks are required to preserve the intended filtering 
behavior, as illustrated in Fig.~\ref{fig:Higher_Order_NPF}. The isolation network can be realized using either active or passive circuits, depending on the design requirements.

\begin{figure}[!t]
  \centering
  \includegraphics[width =0.8\columnwidth]{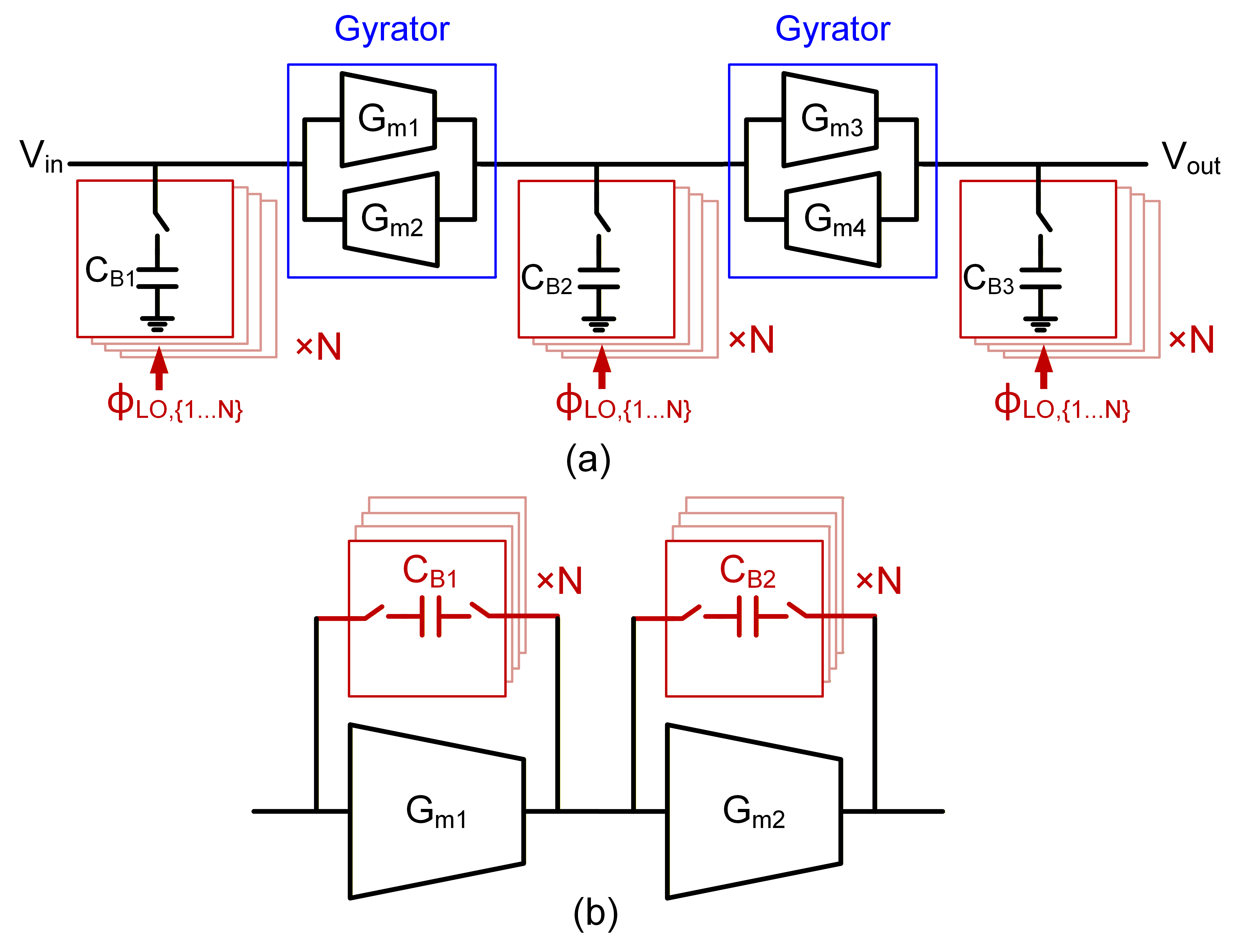}
  \caption{(a) Higher-order N-path filter with active gyrators 
  as isolation networks~\cite{darvishi_jssc_2013}. 
  (b) Cascaded gain-boosted N-path filters~\cite{razavi_jssc_2022}.}
  \label{fig:Higher_Order_NPF_CKT}
\end{figure}

A practical implementation of a higher-order N-path filter is 
shown in Fig.~\ref{fig:Higher_Order_NPF_CKT}(a), where active 
gyrators are used as isolation networks between cascaded N-path sections~\cite{darvishi_jssc_2013}. This circuit realizes a tunable sixth-order N-path channel-select filter. However, the $\mathrm{G_m}$ cells and 
switches introduce substantial parasitic capacitance at the 
internal nodes, which increasingly degrades performance at 
higher frequencies. In FR3, these parasitics introduce 
additional loss and reduce blocker rejection. In addition, the 
noise and nonlinearity of the active isolation circuits degrade the NF and linearity of the RX, particularly in the presence of strong blockers.

A higher-order N-path filter based on cascaded gain-boosted 
N-path sections is shown in Fig.~\ref{fig:Higher_Order_NPF_CKT}(b) \cite{razavi_jssc_2022}. The gain-boosted architecture reduces the required switch and capacitor sizes by approximately a factor of $1+A_0$~\cite{park_jssc_2014}, improving the frequency-scalability of the N-path filter. However, the gain-boosting operation is highly sensitive to parasitic capacitance in the feedback paths, particularly at higher frequencies. The two series switches shown in Fig.~\ref{fig:Higher_Order_NPF_CKT}(b) are introduced to mitigate this effect.

In FR3, the parasitic capacitance of switches, capacitors, and $\mathrm{G_m}$ cells in higher-order N-path filters introduces additional loss and limits the blocker rejection. A potential solution is to use inductors and transformers, which are physically smaller in FR3 than in FR1, to mitigate the effects of parasitic capacitance. However, the higher loss of switches at FR3 bands increases loss and NF for higher-order N-path filters. In addition, cascading multiple N-path filters to enhance selectivity results in higher loss, compared to using higher-order BB impedance. Generally, selectivity enhancement approaches based on higher-order BB impedance, which offer higher resilience to parasitics and losses, are preferred for FR3.

\section{Harmonic Rejection}
\label{section:harmonic}

\begin{figure}[!t]
  \centering
  \includegraphics[width = 0.8\columnwidth]{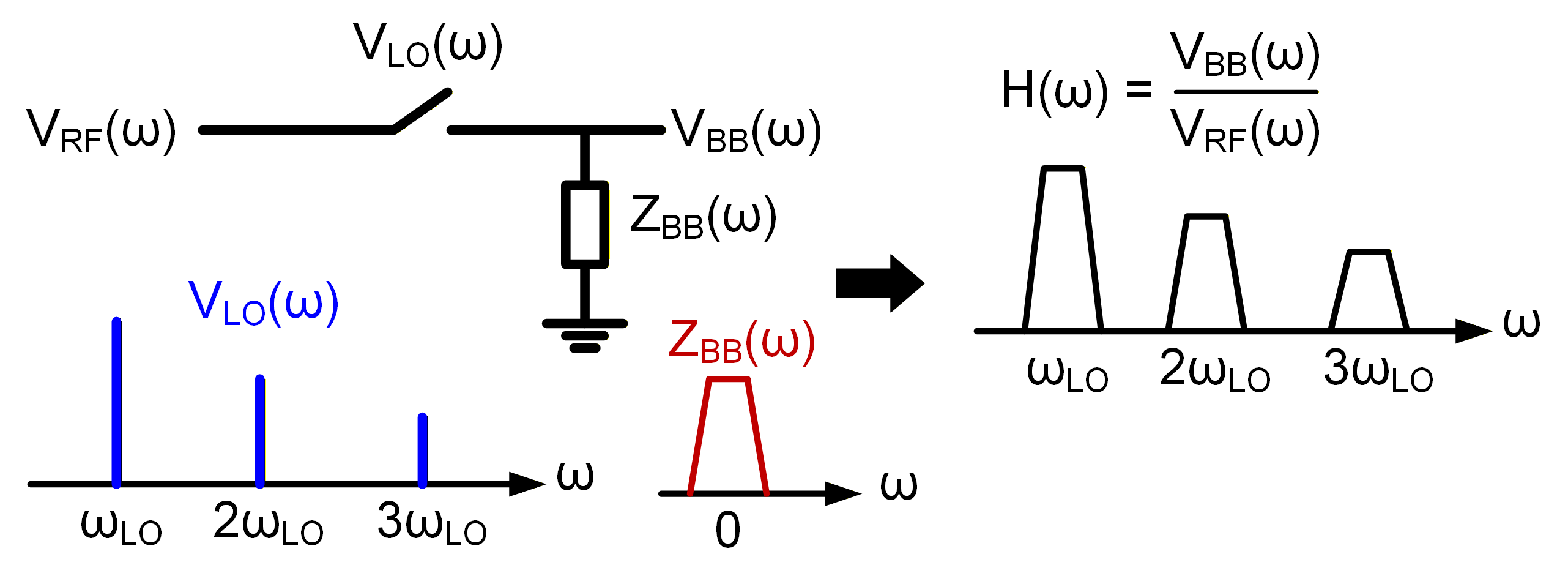}
  \caption{Transfer function of the mixer-first RX with the LO harmonics. The frequency translational property of the switch mixer translates the BB impedance profile to the fundamental and harmonics of the LO frequency.}
  \label{fig:HR_basics}
\end{figure}

Mixer-first RXs usually operate using square-wave LO waveforms that contain harmonics of the LO frequency. Therefore, any blocker at LO harmonics is down-converted to BB, making mixer-first RXs susceptible to harmonic blockers. Even-order harmonics can be suppressed using differential circuits, but dedicated techniques are required to reject odd-order harmonics. As illustrated in Fig.~\ref{fig:HR_basics}, the frequency translational property of the mixer translates the BB impedance profile not only around the fundamental LO frequency, but also around its harmonics. Therefore, blockers located near LO harmonics can be down-converted to BB alongside the desired signal, making harmonic rejection a critical requirement for mixer-first RXs.

\begin{figure}[!t]
  \centering
  \includegraphics[width = 0.5\columnwidth]{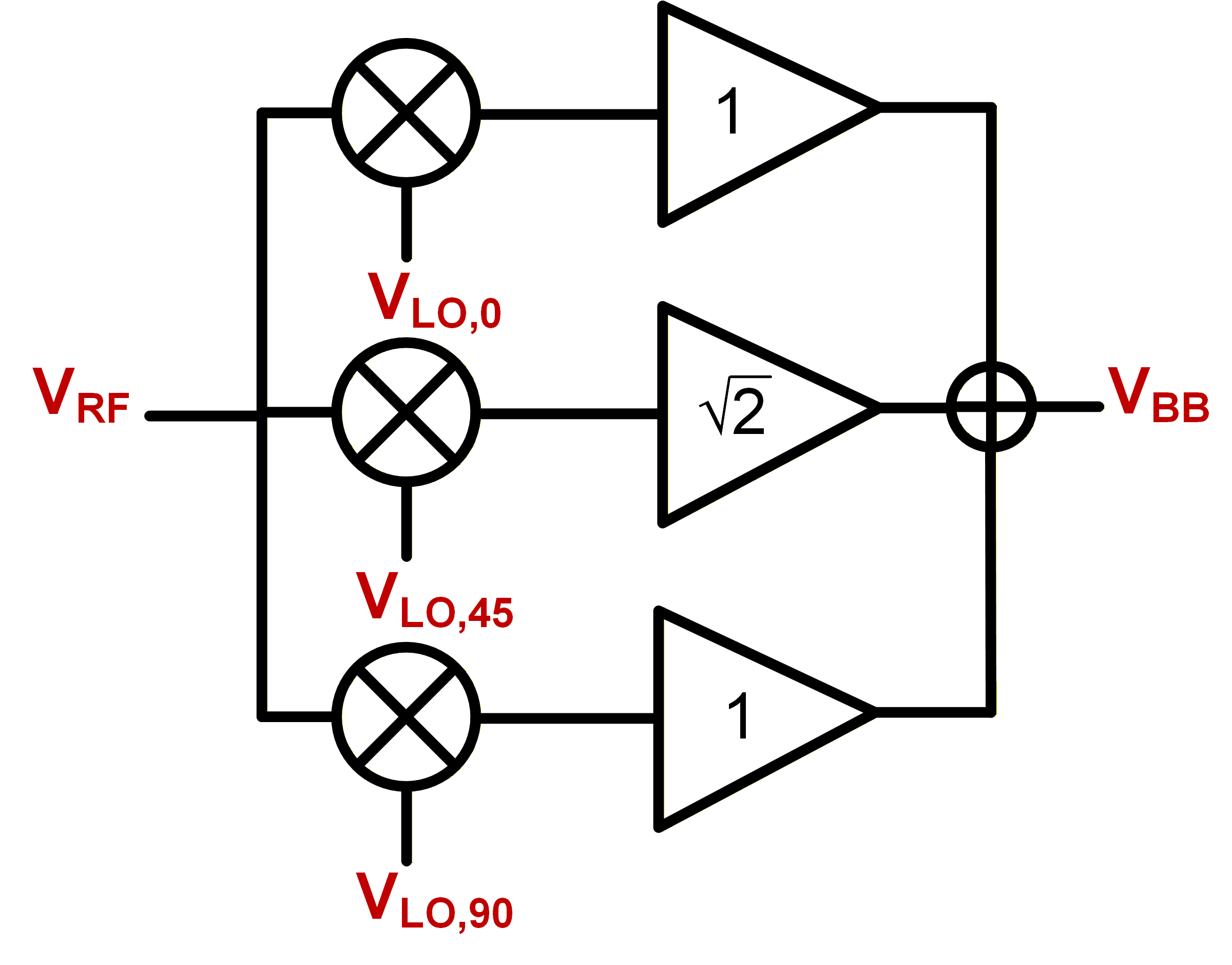}
  \caption{Harmonic rejection mixer with LO phases of $\{0^{\circ}, -45^{\circ}, -90^{\circ}\}$ and relative weights of $\{1, \sqrt{2}, 1\}$. The relative weights can be applied at either BB or RF. Signals at the fundamental LO frequency combine constructively, 
while the third and fifth LO harmonics are ideally canceled.}
  \label{fig:HRM}
\end{figure}

A popular harmonic-rejection mixer (HRM) is shown in 
Fig.~\ref{fig:HRM}. The HRM comprises three sub-mixers driven 
by LO phases of $\{0^{\circ}, -45^{\circ}, -90^{\circ}\}$, 
whose outputs are combined with relative weights of 
$\{1,\sqrt{2},1\}$ \cite{Weldon_jssc_2001}. Assuming ideal 
sub-mixers, the conversion gain at the $n$-th LO harmonic can 
be derived as
\begin{equation}
H(n\omega_0)=\left|\sqrt{2}+2\cos\left(\frac{n\pi}{4}\right)\right|.
\end{equation}
This transfer function produces constructive addition at the fundamental LO frequency while ideally canceling the third and fifth LO harmonics. In practice, amplitude and phase mismatch among the three paths limit the achievable harmonic rejection 
to approximately 30~dB unless calibration or additional RF/BB harmonic rejection techniques are employed.

\subsection{Active Harmonic Rejection}

The factor $\sqrt{2}$ in the HRM circuit must be implemented 
with high precision to achieve strong harmonic rejection. In practice, this factor is usually approximated by a fractional value realized through the size ratio of two on-chip devices, such as transistors or resistors. Some fractional values used in practice include BB transconductance ratios of 16/11 \cite{Andrews_jssc_2010}, 17/12 \cite{xu_jssc_2018}, RF LNA weight ratio of 17/12 \cite{xu_jssc_2018}, RF LNTA weight ratio of 3/2 \cite{ru_jssc_2009}, and BB resistance ratio of 7/5 \cite{ru_jssc_2009}. 

\begin{figure}[!t]
  \centering
  \includegraphics[width = 0.7\columnwidth]{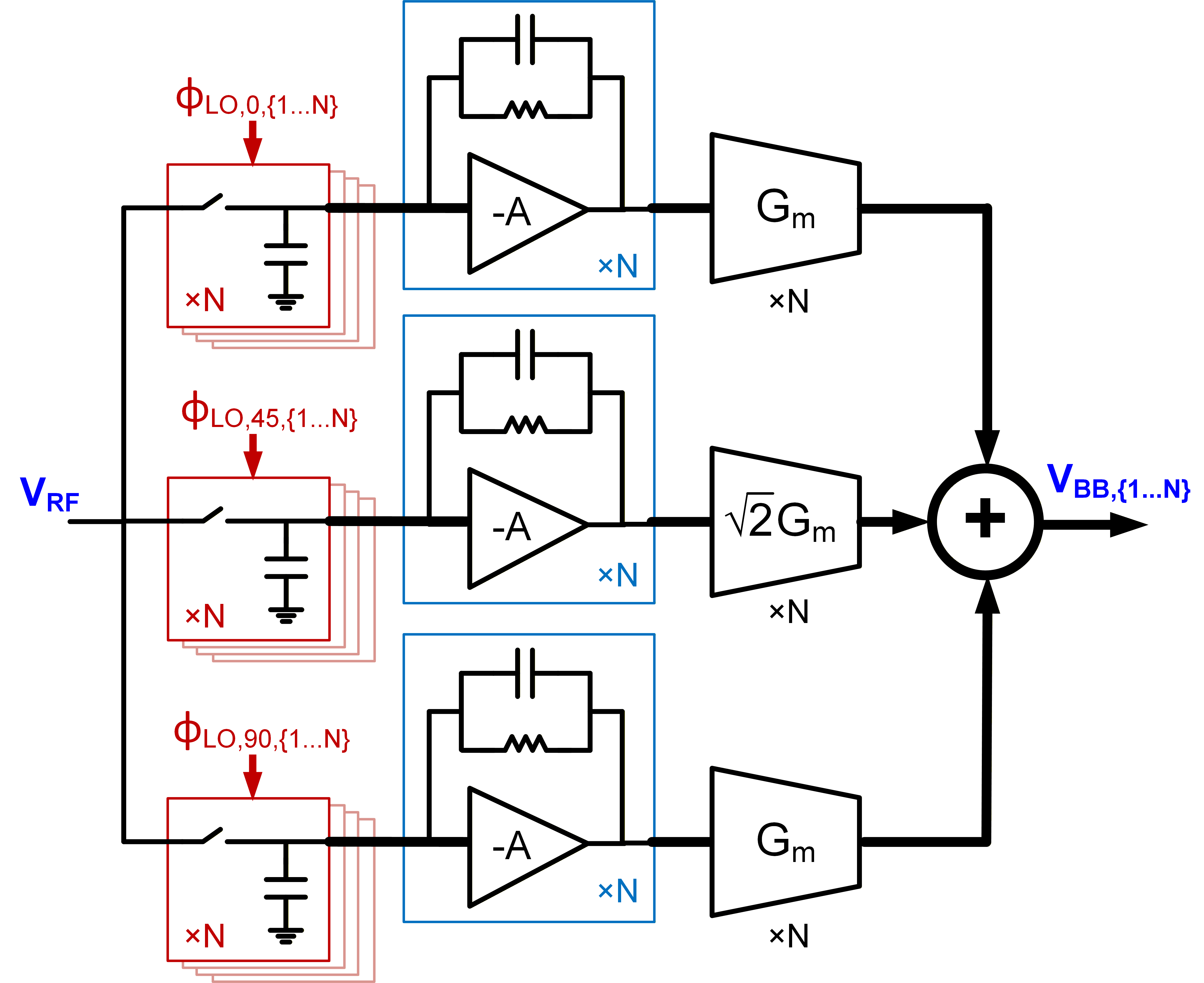}
  \caption{Active harmonic rejection in the mixer-first RX. This general architecture can be used for any number of paths. For an eight-path RX, $N=8$, the three sub-mixers can be merged as their LO phases will be the same. The relative weights should be applied to the BB outputs of the merged sub-mixers.}
  \label{fig:BB_HR}
\end{figure}

In mixer-first RXs, active harmonic rejection is typically implemented using BB harmonic recombination circuits placed after (or before) the BB amplifiers, as shown in  Fig.~\ref{fig:BB_HR}. Early implementations using eight-phase mixers achieved modest $\rm HR_{3,5}$ levels around 35--42~dB \cite{Andrews_jssc_2010}, while more refined designs have pushed rejection to 52--54~dB \cite{Murphy_jssc_2015}, with calibration extending performance beyond 70~dB in some cases \cite{Liempd_jssc_2014}. However, BB  harmonic recombination alone is generally insufficient for most practical applications, and is therefore commonly combined with RF harmonic rejection techniques. These include weighted RF signal paths \cite{ru_jssc_2009, xu_jssc_2018} and cascaded N-path filters with harmonic traps \cite{razavi_jssc_2022}, achieving $\rm HR_{3,5}$ levels of 60--70~dB, which can be further extended beyond 80~dB with digital adaptive interference canceling \cite{ru_jssc_2009}.

These HR approaches rely on active RF front-ends that partially compromise the passive translational operation of mixer-first RXs. Under strong blocker conditions at the RX input, the active circuits can deviate from linear operation and violate the conditions required for effective harmonic rejection. These implementations have been limited to LO frequencies up to 2 GHz, as presented in \cite{razavi_jssc_2022}, and cannot be readily extended to FR3 due to the increased impact of parasitic capacitance at RF nodes. In addition, both BB and RF harmonic rejection circuits require accurate eight-phase clocks with a 12.5\% duty cycle, which becomes increasingly difficult to achieve in FR3 due to phase mismatch and CK jitter.

\subsection{Passive Harmonic Rejection}

\begin{figure}[!t]
  \centering
  \includegraphics[width = 0.9\columnwidth]{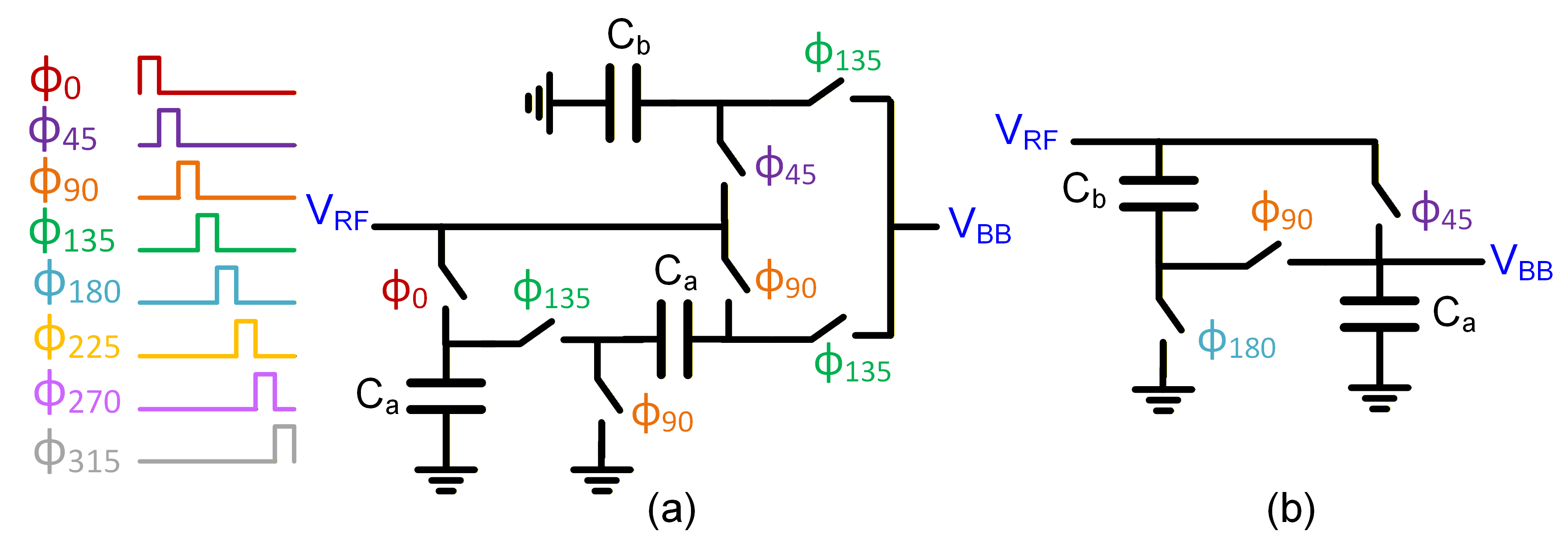}
  \caption{Passive harmonic rejection in the mixer-first RX. (a) Harmonic rejection mixer using charge sharing and capacitor stacking, comprising three capacitors and seven switches \cite{araei_jssc_2023}, (b) Harmonic rejection mixer using bottom and top plate mixing, comprising two capacitors and three switches \cite{araei_jssc_2024}.}
  \label{fig:Passive_HR}
\end{figure}

The weighting required for the HRM circuit can also be realized using passive switched-capacitor techniques \cite{araei_jssc_2026, araei_jssc_2024, Weinreich_jssc_2023, araei_jssc_2023}. These approaches are inherently compatible with mixer-first RX architectures and are potentially more scalable to FR3 because they avoid active RF weighting circuits. The passive HRM shown in Fig.~\ref{fig:Passive_HR}(a) operates 
based on charge sharing and capacitor stacking principles 
\cite{araei_jssc_2023}. The resulting BB voltage can be 
expressed as
\begin{equation}
    \label{SC_HRM}
    V_\mathrm{BB,0^{\circ}} =
    \frac{
    V_\mathrm{RF,0^{\circ}}
    +
    \sqrt{2}V_\mathrm{RF,45^{\circ}}
    +
    V_\mathrm{RF,90^{\circ}}
    }{1+\sqrt{2}},
\end{equation}
which follows the same phase and amplitude relationships as 
the HRM of Fig.~\ref{fig:HRM}. Experimental implementations based on this approach at sub-4 GHz bands have demonstrated $\rm HR_3$/$\rm HR_5$ exceeding 45/50 dB together with strong blocker linearity \cite{araei_jssc_2023}. 
The passive HRM shown in Fig.~\ref{fig:Passive_HR}(b) leverages 
bottom-plate and top-plate mixing to perform harmonic 
rejection at both the RF and BB ports of the mixer 
\cite{araei_jssc_2024}. The resulting BB voltage follows the 
same form as (\ref{SC_HRM}). Compared to the HRM of 
Fig.~\ref{fig:Passive_HR}(a), this implementation requires 
fewer switches and capacitors, making it less sensitive to 
parasitic capacitance and passive losses. Experimental results have demonstrated harmonic rejection levels approaching 50 dB and strong third-harmonic blocker 1-dB compression point (H3-B1dB) of 14 dBm, measured at 1 GHz. 
In Table \ref{tab:HR}, a summary of harmonic rejection performance for mixer-first RXs is presented. This indicates that passive RF harmonic rejection can significantly enhance blocker suppression. However, the performance degrades at higher frequencies, mainly due to switch loss and parasitic capacitance.

Passive HRMs avoid active RF weighting, but their additional switches and capacitors increase loss and input loading. Active BB recombination uses fewer RF sampling elements but requires accurate wideband gain ratios and sufficient blocker handling. Neither family is categorically more scalable. The appropriate choice follows from the required harmonic rejection, phase count, NF, input capacitance, and clock power.

The harmonic requirement must be derived from the complete tuning and blocker-frequency plan. For example, the third harmonic of an LO in the 7.125--8.4-GHz range overlaps 21.375--25.2~GHz, including part of upper FR3. Harmonics outside FR3 cannot be ignored because incumbent transmitters at those frequencies may still reach the antenna and be translated into the channel. The relevant quantity is therefore the available harmonic-blocker power multiplied by the mixer harmonic response, not whether the blocker shares the receiver's nominal allocation.

\begin{table}
        \caption{Harmonic rejection (HR3/HR5) performance and techniques for reported mixer-first RXs.}
        \renewcommand{\arraystretch}{1.25}
        \setlength{\tabcolsep}{2pt}
         \centering
    \begin{tabular}{ccccc}
    \hline
       Ref  & f (GHz)  & HR (dB) & HR Techniques  & Process\\
       \hline
       \hline
         \cite{Andrews_jssc_2010} & 0.5  & 35/42 & Active BB & 65-nm CMOS \\
       \cite{araei_jssc_2023}  & 0.5 & 55/52  &  Passive RF, Active BB & 45-nm PDSOI\\
        \cite{araei_jssc_2023} & 2.5  & 43/47 & Passive RF, Active BB & 45-nm PDSOI\\
         \cite{araei_jssc_2024} & 1.0  & 48/54 & Passive RF, Active BB & 45-nm PDSOI\\
         \cite{Weinreich_jssc_2023}& 1.0 & 56/54 &  Passive RF, Active BB & 22-nm FDSOI\\
         \cite{Murphy_jssc_2015}& 2.0 & 60/60 & RF NC, BB HR-TIA & 28-nm CMOS\\
         \hline
    \end{tabular}
    \label{tab:HR}
\end{table}

\section{Linearization Techniques}
\label{section:linearity}

The main sources of nonlinearity in mixer-first RXs are the nonlinearity of the switch on-state resistance and the nonlinearity of the BB amplifiers. These nonlinearities significantly affect blocker tolerance and become increasingly challenging in FR3 due to reduced voltage headroom in advanced CMOS technology nodes, higher CK frequencies, and increased RF parasitic sensitivity. 
The most effective linearization techniques for mixer-first RXs are discussed in this section, including bottom-plate mixing, switch CK boosting and bootstrapping, and BB amplifier linearization.

\subsection{Bottom-Plate Mixing}

\begin{figure}[!t]
  \centering
  \includegraphics[width = 0.9\columnwidth]{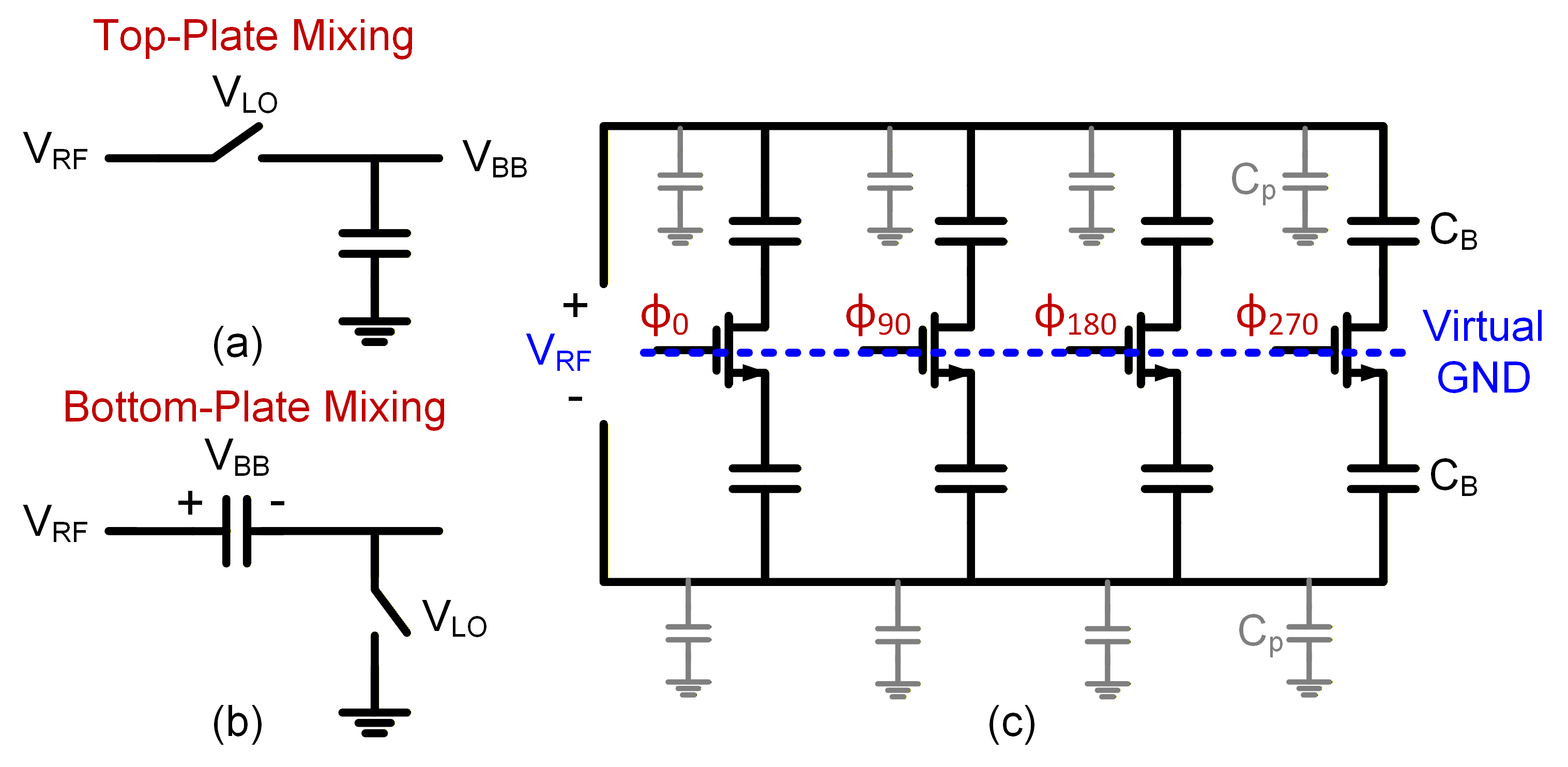}
  \caption{(a) Top-plate mixing, (b) bottom-plate mixing, (c) a mixer-first RX using bottom-plate mixing. The parasitic capacitance of the capacitors, shown by $C_p$, accumulates at the RF input node and increases the loss at higher frequencies.}
  \label{fig:Bottom_Plate_Mixing}
\end{figure}

In conventional top-plate mixing, shown in Fig. \ref{fig:Bottom_Plate_Mixing}(a), the switch is placed between the RF signal and the BB capacitor. In this structure, the on-state gate-source voltage of the switch depends on the input voltage, $V_{GS} = V_{DD} - V_{RF}$, which increases the nonlinearity of the switch on-resistance. In bottom-plate mixing, shown in Fig. \ref{fig:Bottom_Plate_Mixing}(b), the switch is connected to the bottom plate of the capacitor. This results in a constant on-state gate-source voltage, $V_{GS} = V_{DD}$, improving the linearity of the switch on-resistance and the overall sampling circuit.

Bottom-plate mixing can be applied to mixer-first RXs, as shown in Fig.~\ref{fig:Bottom_Plate_Mixing}(c). Operating across 0.1--2.0 GHz, the RX featuring this technique \cite{Lien_jssc_2019} demonstrated improved IB and OOB Input Third-Order Intercept Point (IIP3) by 10 dB and 6 dB, respectively. However, the parasitic capacitance of the BB capacitors accumulates at the RF input node, increasing RF loss and degrading high-frequency performance. This issue becomes more severe in FR3, where RF parasitics and passive losses strongly affect RX performance. Although an input transformer can partially resonate out the parasitic capacitance, the limited Q factor of on-chip transformers may reduce the achievable benefit.

\subsection{Clock Boosting for Switches}

\begin{figure}[!t]
  \centering
  \includegraphics[width = 0.9\columnwidth]{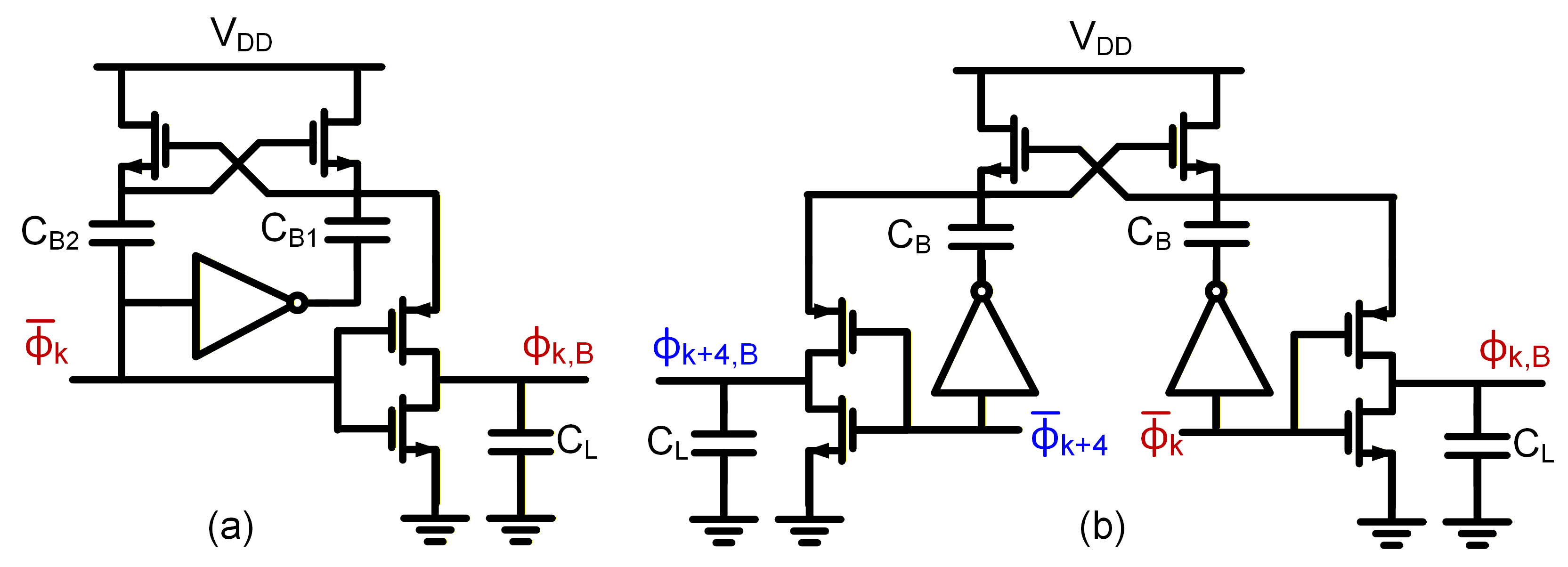}
  \caption{Clock boosting for switches. (a) single-phase charge pump, (b) charge pump with complementary phases (used for an eight-phase clock). The circuits can ideally provide the boosting factor of 2.}
  \label{fig:CK_Boosting}
\end{figure}

The main idea of CK boosting is to increase the amplitude of the CK signal to mitigate the impact of switch on-resistance nonlinearity. A popular charge pump CK boosting circuit is shown in Fig. \ref{fig:CK_Boosting}(a) \cite{Nakagome_jssc_1991, Cho_jssc_1995}. In the gain-boosted N-path RX presented in \cite{araei_jssc_2026}, a modified CK boosting circuit with complementary phases is introduced to halve the required number of CK paths in the RX, as shown in Fig. \ref{fig:CK_Boosting}(b). However, the available time to recharge the boosting capacitor $C_B$ is reduced from $\frac{7}{8} T_{LO}$ to $\frac{1}{8} T_{LO}$, i.e., by a factor of 7, which can limit the achievable boosting factor at high LO frequencies. The boosting factor is 1.9 at 0.1 GHz and decreases to 1.5 at 3 GHz. Improving the boosting factor at higher frequencies requires increased power consumption.

In FR3, the higher operating frequencies require high-speed charge-pump CK boosting circuits to achieve large boosting factors, ideally close to 2. However, the reduced time available for recharging the boosting capacitors at high LO frequencies can limit the achievable boosting factor, increase power consumption, and ultimately constrain the frequency scalability of CK boosting in FR3. At the same time, CK boosting becomes increasingly important in advanced CMOS technologies due to reduced supply voltage and increased sensitivity to switch nonlinearity.

\subsection{Clock Bootstrapping for Switches}

\begin{figure}[!t]
  \centering
  \includegraphics[width = 0.6\columnwidth]{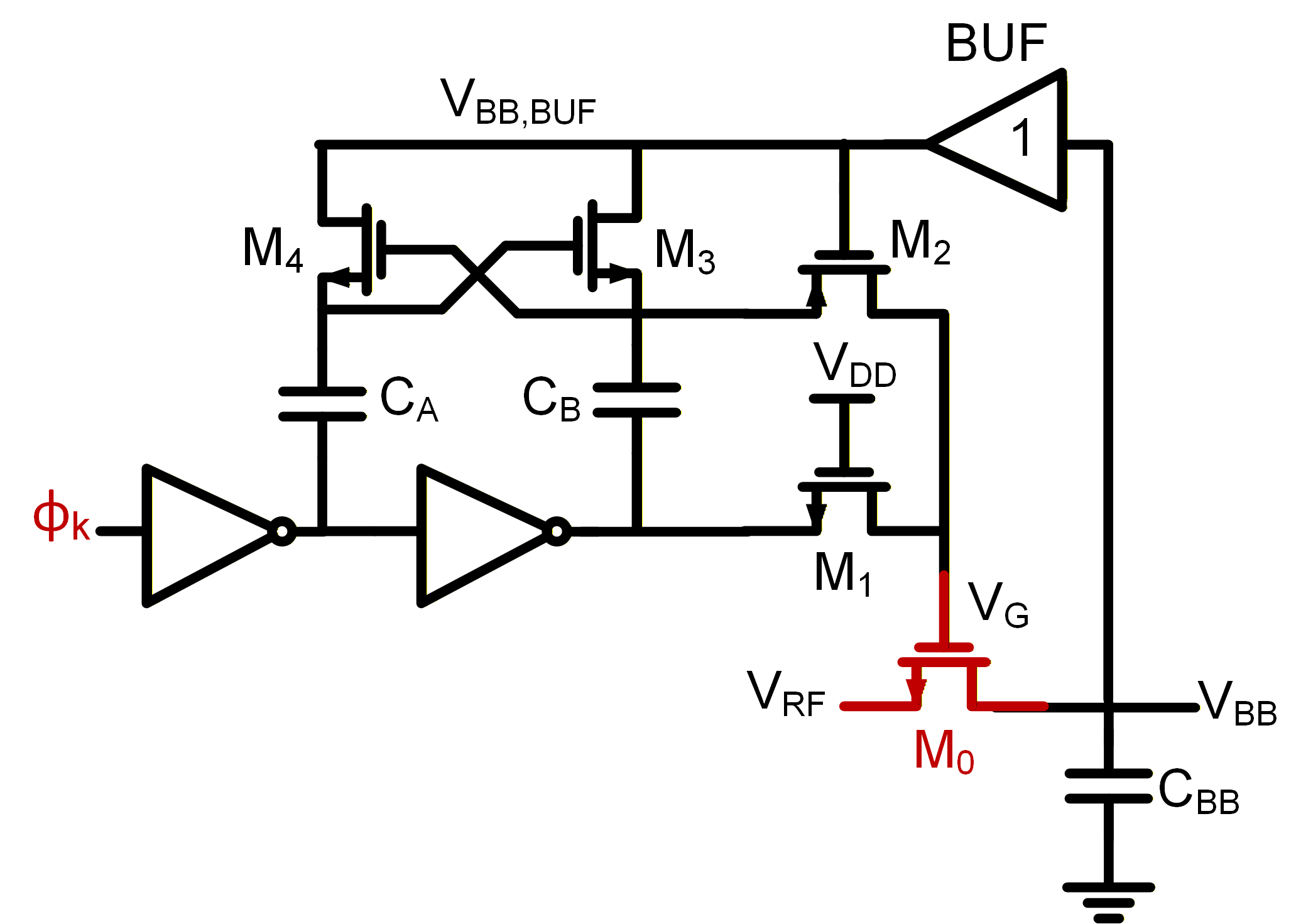}
  \caption{Clock bootstrapping circuit for the switch transistor $M_0$ \cite{Hardeveld_jssc_2026}. The BB voltage $V_{BB}$ is stored on the capacitors $C_A$ and $C_B$ when the clock is low. The stored voltage is used to generate a constant voltage of $V_{DD}$ across the gate-source of the switch transistor when the clock is high.}
  \label{fig:CK_Bootstrapping}
\end{figure}

The main idea of CK bootstrapping is to buffer and store the input voltage during the switch off-state, and use it as a reference to generate a constant gate-source voltage for the switch during its on-state \cite{Dessouky_EL_1999, Abo_jssc_1999, Hardeveld_jssc_2026}. A CK bootstrapping circuit proposed for a mixer-first RX is shown in Fig. \ref{fig:CK_Bootstrapping} \cite{Hardeveld_jssc_2026}. When the CK is low, the gate of the switch transistor $M_0$ is connected to ground through $M_1$, thus $V_G=0$. The voltage $V_{BB}$ is buffered and stored on the capacitors $C_A$ and $C_B$. When the CK is high, the switch gate is connected to the top plate of $C_B$ through $M_2$, thus $V_G = V_{BB} + V_{DD}$ and $V_{GS,M0} = V_{DD}$. The mixer-first RX achieves excellent IB-IIP3 of 12--18 dBm in the 1--8 GHz band. 

The results presented in \cite{Hardeveld_jssc_2026} indicate that this approach may be extendable toward FR3 operation. However, such an extension would require the bootstrapping circuit to operate at very high clock speeds, increasing both implementation complexity and power consumption.

\subsection{Linearized BB Amplifier}

In a mixer-first RX, the BB impedance is translated to RF by the periodically time-varying passive mixer. Therefore, the BB amplifier affects not only post-mixer gain, but also the RF input impedance, blocker tolerance, linearity, and NF of the RX \cite{Borremans_jssc_2011, lin_jssc_2014}. This makes BB-domain linearization attractive, since improvements in BB impedance linearity can be reflected at the RF input.

In a conventional mixer-first RX, the finite input impedance of the BB amplifier can contribute to RF input matching after impedance translation. However, large IB signals or blockers can create a non-negligible voltage swing at the BB input node. Voltage-dependent variations of this impedance are then translated to RF and can appear as nonlinear input impedance, producing distortion \cite{ru_jssc_2009, Borremans_jssc_2011}.

\begin{figure}[!t]
  \centering
  \includegraphics[width = 0.75\columnwidth]{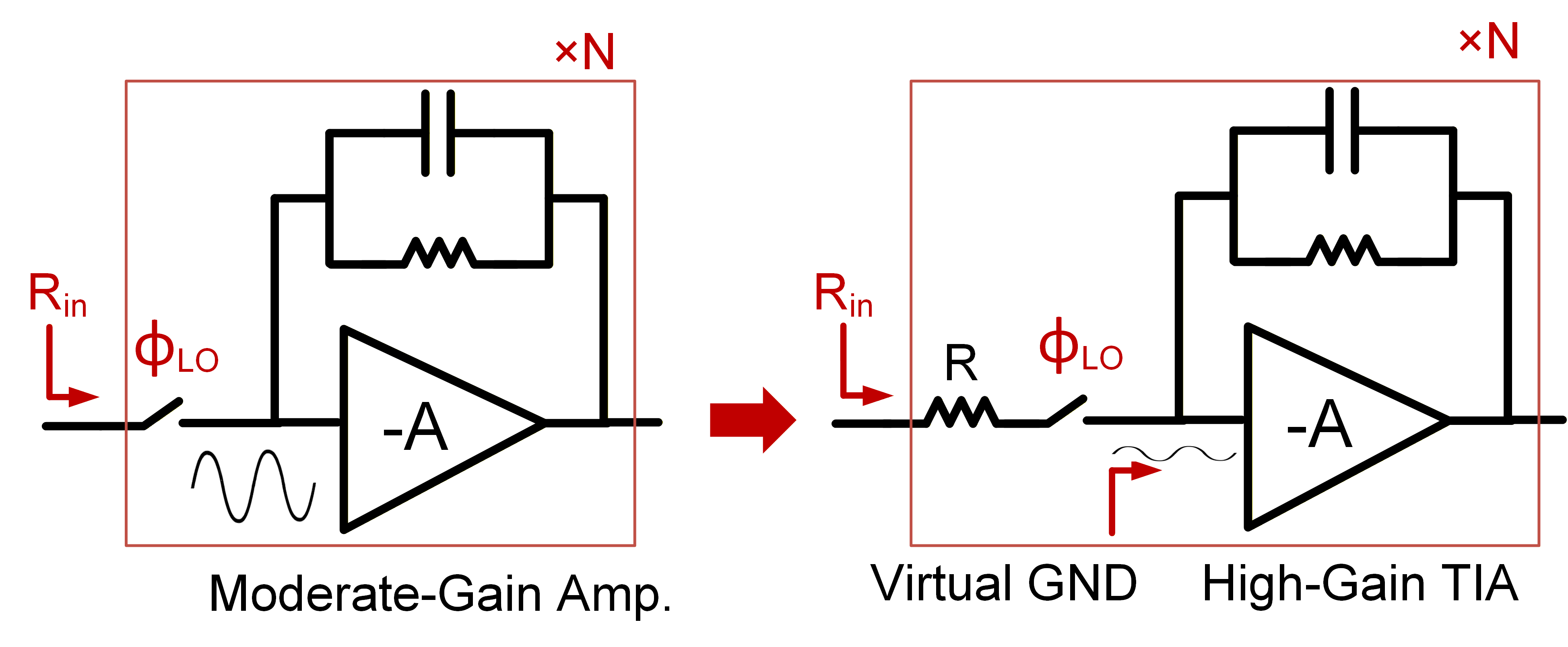}
  \caption{BB amplifier linearization with virtual-ground input. A moderate-gain amplifier can provide input matching, but the voltage swing at the input of the amplifier degrades linearity. A high-gain TIA can provide a virtual ground input with very small input voltage swing to improve linearity. A series resistance should be used for input matching.}
  \label{fig:BB_AMP_linearization}
\end{figure}

A linearized alternative is shown in Fig.~\ref{fig:BB_AMP_linearization}. The BB amplifier is configured with high loop gain so that its input behaves approximately as a virtual ground. This suppresses the BB input voltage swing and relaxes the linearity requirement of the amplifier input devices. Since the low-impedance condition is translated to RF, the RX can maintain a more linear input termination around the LO frequency. Because the virtual-ground input provides little real resistance for matching, an additional passive resistance is introduced to help satisfy the input matching condition \cite{Krishnamurthy_jssc_2021}. Experimental implementations using this principle have demonstrated high IB linearity over the 10--35~GHz range \cite{Krishnamurthy_jssc_2021}.

The main drawback is noise. In a true mixer-first RX, with little or no RF gain before the mixer, the thermal noise of the added resistance is not suppressed by preceding gain and can directly degrade NF. This is consistent with the general mixer-first tradeoff: removing RF gain improves blocker tolerance and linearity, but makes the mixer switches, LO generation, and first BB stage more critical to noise \cite{Andrews_jssc_2010, Andrews_jssc_2013, Borremans_jssc_2011, Lin_jssc_2014_2}. A possible compromise is to realize only part of the required input resistance with an explicit passive resistor and synthesize the remaining part using the finite input impedance of the BB amplifier. This reduces resistor noise, but increases the BB input swing and can degrade linearity.

This tradeoff is architecture-dependent. If the N-path mixer is driven directly from the antenna, the added resistance appears before the first significant gain stage and its NF penalty is severe. In LNTA-assisted or current-mode N-path RXs, a preceding LNTA can provide current gain and may reduce the need for explicit matching resistance at the mixer interface. Nevertheless, any resistance remaining in the signal-current path still injects thermal noise into the mixer/TIA input. This distinction is related to current-mode versus voltage-mode N-path operation: in current mode, a BB virtual ground is translated to RF, whereas in voltage mode, a BB low-pass RC impedance is translated to RF to form an N-path bandpass response \cite{Lin_jssc_2014_2, Andrews_jssc_2013}.

Finally, virtual-ground operation suppresses one important distortion mechanism, but does not remove all nonlinearities. After the BB input swing is reduced, the dominant distortion may shift to the TIA feedback impedance, finite gain-bandwidth, output swing, or blocker-current handling capability. Therefore, BB amplifier linearization should be co-designed with input matching, selectivity, harmonic rejection, and noise optimization \cite{ru_jssc_2009, Lin_jssc_2014_2}. Other BB linearization and feedback techniques \cite{Zhang_tcas1_2011, Subramaniyan_jssc_2015}, as well as lower-frequency noise-canceling approaches \cite{Murphy_jssc_2012, Murphy_jssc_2015, Bhat_jssc_2021}, may also be useful if their bandwidth, phase accuracy, auxiliary-path linearity, and parasitic sensitivity remain acceptable.

\subsection{Discussion}

\begin{figure}[!t]
  \centering
  \includegraphics[width = 0.8\columnwidth]{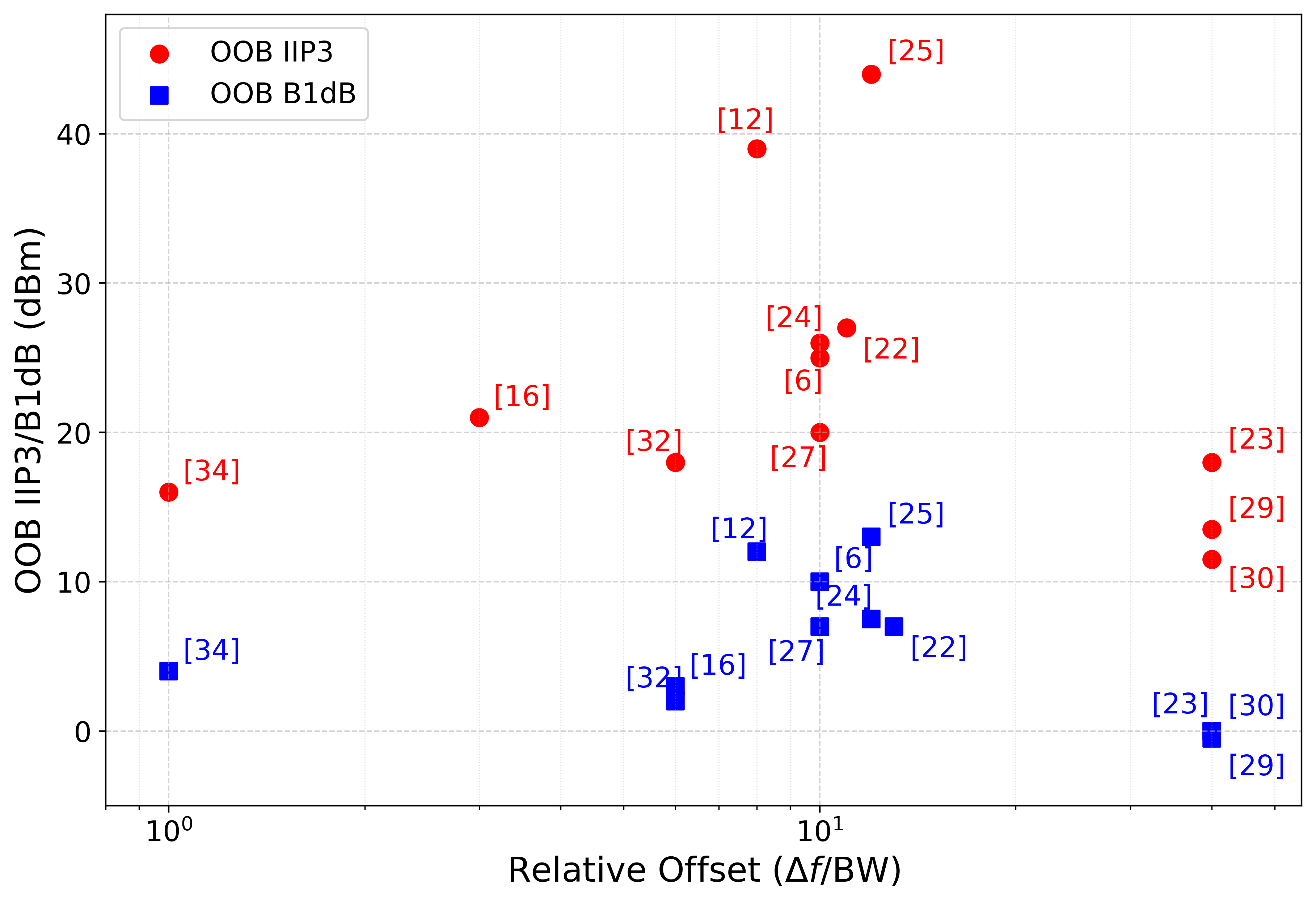}
  \caption{OOB IIP3/B1dB versus relative offset $\rm \Delta f/BW$ for reported mixer-first RXs.}
  \label{fig:OOB_IIP3_B1dB_relative_offset}
\end{figure}

Linearity of a mixer-first RX should be evaluated in the presence of blockers which, as discussed in Section \ref{section:fundamentals} can be categorized as IB, OOB, and harmonic blockers. This implies that linearity metrics should be presented under three conditions, e.g., three blocker 1-dB compression point (B1dB) metrics as IB-B1dB, OOB-B1dB, and H-B1dB. However, in most of the work presented in the literature, the linearity metrics have been reported only for limited types of blockers. In Table \ref{tab:IIP3_B1dB}, a summary of IIP3 and B1dB for reported mixer-first RXs is presented, where it is observed that IB and H linearity metrics have not been reported for many designs. Additionally, OOB linearity metrics are reported at different relative frequency offsets $\rm \Delta f/BW$, as shown in Fig. \ref{fig:OOB_IIP3_B1dB_relative_offset}, which precludes a fair comparison. A potential solution is to develop a standard for reporting OOB linearity metrics, e.g., the adjacent channel $\Delta f/BW = 1$, the alternate channel $\Delta f/BW = 2$, and a far-out channel $\Delta f/BW = 10$.

Linearity of a mixer-first RX is often interrelated with other properties of the RX. In particular, selectivity enhancement usually provides extra suppression of OOB and harmonic blockers, while harmonic rejection improves harmonic blocker linearity. In addition, increasing the switch size to reduce NF also mitigates the impact of switch nonlinearity on the overall RX linearity. Therefore, it is not always necessary to use a dedicated linearization technique to achieve a decent OOB linearity, which can also be inferred from Table \ref{tab:IIP3_B1dB}. Specifically, OOB-IIP3 of 39 dBm and OOB-B1dB of 12 dBm in \cite{Lien_jssc_2018} were achieved using a selectivity enhancement technique. Using bottom-plate mixing, an OOB-IIP3 of 44 dBm is presented in \cite{Lien_jssc_2019}. These are illustrated in Fig. \ref{fig:OOB_IIP3_B1dB_relative_offset}. In addition, the best-in-class H3-B1dB values of 10 dBm in \cite{araei_jssc_2023} and 14 dBm in \cite{araei_jssc_2024} were achieved using passive harmonic rejection circuits (shown in Fig. \ref{fig:Passive_HR}).

In contrast, IB linearity usually requires dedicated linearization techniques. In Table \ref{tab:IIP3_B1dB}, an IB-IIP3 of 15 dBm is reported in \cite{Lien_jssc_2019} using bottom-plate mixing, while \cite{Hardeveld_jssc_2026} reports an IB-IIP3 of 16 dBm by leveraging a combination of CK bootstrapping and BB amplifier linearization.

Experimental results indicate that the highest reported IB linearity levels in mixer-first RXs are achieved using techniques such as bottom-plate mixing, CK bootstrapping, and BB amplifier linearization. Among these approaches, BB-domain linearization techniques are particularly promising for FR3 because they leverage the frequency translational property of mixer-first RXs and are less sensitive to RF parasitics and passive losses.

\begin{table*}
        \caption{Summary of IIP3/B1dB for reported mixer-first RXs. For OOB IIP3/B1dB, relative offset $\rm \Delta f/BW$ is presented in parenthesis.}
        \renewcommand{\arraystretch}{1.25}
        \setlength{\tabcolsep}{4pt}
         \centering
    \begin{tabular}{ccccccccccc}
    \hline
       Ref  & f   & BW  & IB-IIP3 & IB-B1dB  & OOB-IIP3  & OOB-B1dB  & H3-IIP3  & H3-B1dB  & Linearization & Process\\
          & (GHz)  & (MHz) & (dBm) &  (dBm) &  (dBm) &  (dBm) & (dBm) &  (dBm) & & \\
       \hline
       \hline
         \cite{Andrews_jssc_2010} & 1.2 & 10 & NR & NR&  25 (10) & 10 (10) & NR & NR & NR & 65-nm CMOS \\
         \cite{Lien_jssc_2018} & 2.0 & 10 & NR & NR &  39 (8) & 12 (8) & NR & NR & NR& 45-nm PDSOI \\
        \cite{Lien_jssc_2019} & 1.0 & 6.5 & 15 & $-$10 &  44 (12)  & 13 (12) & NR & NR & Bottom-Plate Mixing& 28-nm CMOS \\
         \cite{araei_jssc_2023} & 1.0 & 10  & $-$8 & NR & 27 (11) & 7 (13) & NR & 10 & NR& 45-nm PDSOI\\
         \cite{araei_jssc_2024} & 1.0  & 25  & 0 & $-$25 & 26 (10) & 7.5 (12) & 52 & 14 & NR & 45-nm PDSOI\\
         \cite{Weinreich_jssc_2023}& 1.0  & 1 & $-$11 &  $-$33  & 18 (40) & 0 (40) & 26 & $-$8 & NR& 22-nm FDSOI\\
         \cite{Murphy_jssc_2012}& 2.0  &  2 & NR & NR & 13.5 (40)& $-$0.5 (40)& NR & NR & RF Noise-Cancellation & 40-nm CMOS\\
         \cite{Murphy_jssc_2015}& 2.0  & 0.2  & NR & NR  & 11.5 (40) & 0 (40) & NR & $-$6.5 & RF Noise-Canc., HR-TIA& 28-nm CMOS\\
         \cite{Bhat_jssc_2021} & 3.0 & 175 & 9 & $-$11 & 18 (6) & 2 (6) & NR & NR & BB Noise-Cancellation & 22-nm FDSOI \\
         \cite{Hardeveld_jssc_2026} & 4.0 & 190 & 16 & $-$10& 20 (10) & 7 (10) & NR & NR & CK BS, BBA Lin& 22-nm FDSOI \\
        \cite{Hardeveld_jssc_2026} & 8.0 & 190 & 18 & 5 & NR & NR & NR & NR & CK BS, BBA Lin& 22-nm FDSOI \\
         \cite{Wu_tmtt_2016} & 2.0 & 50 & 6 & NR& 16 (1) & 4 (1) & NR & NR & NR & 28-nm CMOS \\
         \cite{Pini_jssc_2020} & 2.0 & 260 & $-$12 & NR& 21 (3) & 3 (6) & NR & NR & NR& 28-nm CMOS \\
         \hline
    \end{tabular}
    \label{tab:IIP3_B1dB}
\end{table*}


\section{Low-Noise Architectures}
\label{section:noise}

\subsection{Noise in Mixer-First Receiver}

In a traditional LNA-first RX, the LNA offers two major advantages: by providing high gain, it suppresses the noise contributions of subsequent RX blocks, and by providing low NF, it dominates the overall NF of the RX. In contrast, a mixer-first RX lacks a front-end gain and this allows multiple circuit blocks contribute to NF. The main noise sources in a mixer-first RX include the switch on-resistance, the feedback resistance of the BB amplifier, the shunt impedance modeling losses due to LO harmonics, and the BB amplifier \cite{Andrews_jssc_2010, Andrews_tcas1_2010, yang_tcas1_2015}.

It is shown that \cite{Andrews_tcas1_2010, yang_tcas1_2015} noise factor of an N-path mixer-first RX can be expressed as
\begin{equation}
\label{NF_MFRX}
    F = 1 + \frac{R_{sw}}{R_s} + \frac{(R_s+R_{sw})^2}{R_s R_{sh}} + \frac{(R_s + R_{sw})^2}{\gamma_N R_s R_F} + k \frac{R_{n}}{R_s},
    \end{equation}
\begin{equation}
    \label{beta}
    k =  \gamma_N  \left( \frac{R_s + R_{sw}}{\gamma_N R_F} + \frac{R_s + R_{sw} + R_{sh}}{R_{sh}}\right)^2,
\end{equation}
where $R_{sh}$ is the resistance associated with harmonic losses and $R_n$ is the input-referred noise resistance of the BB amplifier. A lower NF can potentially be achieved by increasing the switch size to reduce $R_{sw}$, the feedback resistance $R_F$, and the transconductance of the BB amplifier to reduce $R_n$. However, all of these design choices involve trade-offs with other RX performance metrics. Using larger switches increases parasitic capacitance, leading to additional loss and higher NF at higher frequencies of FR3. Larger switches also require greater power consumption in the LO driver circuits. The feedback resistance cannot be increased arbitrarily because the input impedance matching condition must be maintained. Finally, increasing the BB amplifier transconductance also results in higher power consumption.

A blocker-tolerant RX should maintain a low NF even in the presence of strong blockers. Blocker NF (BNF) defined as the NF measured at a certain blocker power (often 0 dBm) and frequency offset is a measure of NF resilience to blockers, and is closely related to noise, linearity, and selectivity. Therefore, BNF is a metric with more practical value than NF for mixer-first RXs. However, many reported mixer-first RXs either do not report BNF \cite{Andrews_jssc_2010, Bhat_jssc_2021, Iotti_jssc_2020, Anton_jssc_2026} or characterize it only at a single frequency and blocker offset.

Table~\ref{tab:NF} summarizes the reported NF and BNF of representative mixer-first RXs. Reported BNF data in the literature remains limited and is often characterized only at 
a single frequency and blocker offset. Existing results indicate that mixer-first RXs employing noise-canceling architectures can achieve sub-2~dB NF with relatively 
small BNF degradation under strong blockers, e.g., approximately 2~dB at a 0-dBm blocker under the reported test conditions \cite{Murphy_jssc_2012, Murphy_jssc_2015}. In addition, recent implementations in advanced CMOS technologies have achieved NF around 6~dB at 8~GHz within the FR3 band \cite{Hardeveld_jssc_2026}. Direct comparison requires the blocker power, offset, and bandwidth listed with each result.

\begin{table}
        \caption{Summary of NF and BNF for reported mixer-first RXs.}
        \renewcommand{\arraystretch}{1.25}
        \setlength{\tabcolsep}{3pt}
         \centering
    \begin{tabular}{cccccc}
    \hline
       Ref  & f (GHz)  & NF (dB) & BNF (dB) & $\rm \Delta f/BW$ & Process\\
       \hline
       \hline
         \cite{Andrews_jssc_2010} & 0.1--2.4 & 3.0--4.0 & NR & NR& 65-nm CMOS \\
         \cite{Lien_jssc_2018} & 0.5--6.0 & 2.3--5.4 & 4.7 & 8 & 45-nm PDSOI \\
         \cite{araei_jssc_2023} & 0.25--2.5 & 3.5--5.5  & 7.8 & 12 & 45-nm PDSOI\\
         \cite{araei_jssc_2024} & 0.25--4.0  & 2.6--5.5  & 9 & 8 & 45-nm PDSOI\\
         \cite{Weinreich_jssc_2023}& 0.3--3.0  & 3.4--4.8 & 13 & 60  & 22-nm FDSOI\\
         \cite{Murphy_jssc_2012}& 0.08--2.7  &  1.5--2.0 & 4.1 & NR & 40-nm CMOS\\
         \cite{Murphy_jssc_2015}& 0.3--3.3  & 1.7--2.2  & 5 & 26  & 28-nm CMOS\\
         \cite{Bhat_jssc_2021} & 1.0--6.0 & 2.5--5.0 & NR & NR & 22-nm FDSOI \\
         \cite{Hardeveld_jssc_2026} & 1.0--8.0 & 4.0--6.0 & 8.8 & 5& 22-nm FDSOI \\
         \cite{Wu_tmtt_2016} & 0.4--3.5 & 2.4--2.6 & 6.5 & 1& 28-nm CMOS \\
         \hline
    \end{tabular}
    \label{tab:NF}
\end{table}

\subsection{Noise-Canceling Receivers}

\begin{figure}[!t]
  \centering
  \includegraphics[width = 0.9\columnwidth]{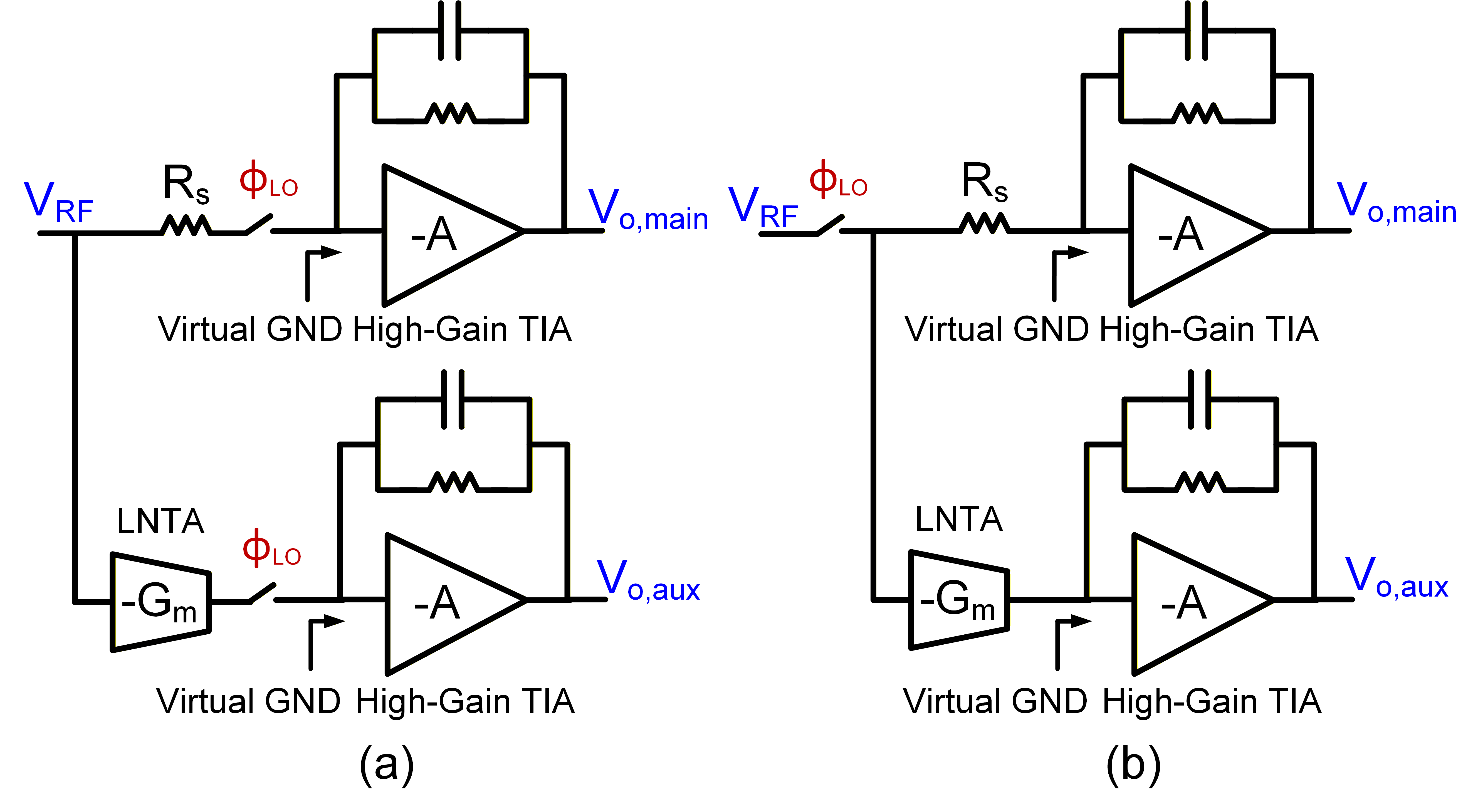}
  \caption{Noise-canceling mixer-first RX architectures. (a) RF noise cancellation, with separate mixers in the main and auxiliary paths, (b) BB noise cancellation with a shared mixer between the main and auxiliary paths.}
  \label{fig:NC_RX}
\end{figure}

A noise-canceling mixer-first RX, shown in Fig. \ref{fig:NC_RX}, comprises a main path that provides input impedance matching through a series resistor $R_s$ and an auxiliary path that cancels the noise contribution of the series resistor. The auxiliary path generates a replica of the series resistor noise voltage at its output, allowing this noise to be canceled by subtracting the outputs of the main and auxiliary paths. Noise cancellation can be implemented either before or after frequency translation, leading to RF- and BB-domain noise-canceling mixer-first RX architectures.

RF and BB noise-canceling architectures exhibit fundamentally different frequency-scaling behaviors for mixer-first RXs in FR3. In RF noise cancellation \cite{Murphy_jssc_2012, Murphy_jssc_2015}, shown in Fig.~\ref{fig:NC_RX}(a), the auxiliary cancellation path operates directly at RF using separate mixers in the main and auxiliary paths. This approach has demonstrated NF below 2~dB over multi-octave bandwidths with BNF of 4--5~dB \cite{Murphy_jssc_2012, Murphy_jssc_2015}. In addition, harmonic rejection was improved from $\rm HR_3$/$\rm HR_5$ of 40~dB in \cite{Murphy_jssc_2012} to 60~dB in \cite{Murphy_jssc_2015} with only modest degradation in BNF, indicating that RF noise cancellation can be effectively combined with harmonic rejection techniques. However, as the LNTA operates at RF, its input parasitic capacitance can degrade input impedance matching and cancellation accuracy, resulting in higher loss and NF in FR3 bands.

In contrast, BB noise cancellation \cite{Bhat_jssc_2021} shown in Fig.~\ref{fig:NC_RX}(b) relocates the auxiliary path to BB while sharing the mixer between the main and auxiliary paths. The LNTA input capacitance then appears at the BB mixer port rather than directly at the RF input, and BB feedback can improve LNTA linearity using established circuit techniques \cite{Zhang_tcas1_2011, Subramaniyan_jssc_2015}. Strong blockers can nevertheless compress the LNTA and perturb the cancellation ratio. The reported implementation achieves 2.5--5-dB NF across 1--6~GHz, but does not report BNF \cite{Bhat_jssc_2021}.

BB noise cancellation reduces direct RF loading by the auxiliary amplifier; RF noise cancellation has demonstrated lower NF and measured BNF in published lower-frequency receivers. Their relative merit in FR3 therefore remains an experimental question. Both NF and cancellation error should be measured versus blocker power and offset.

\section{Multi-Phase Clock Generation}
\label{section:LO}

\subsection{Non-Overlapping CK Generation}

\begin{figure}[!t]
  \centering
  \includegraphics[width = 0.9\columnwidth]{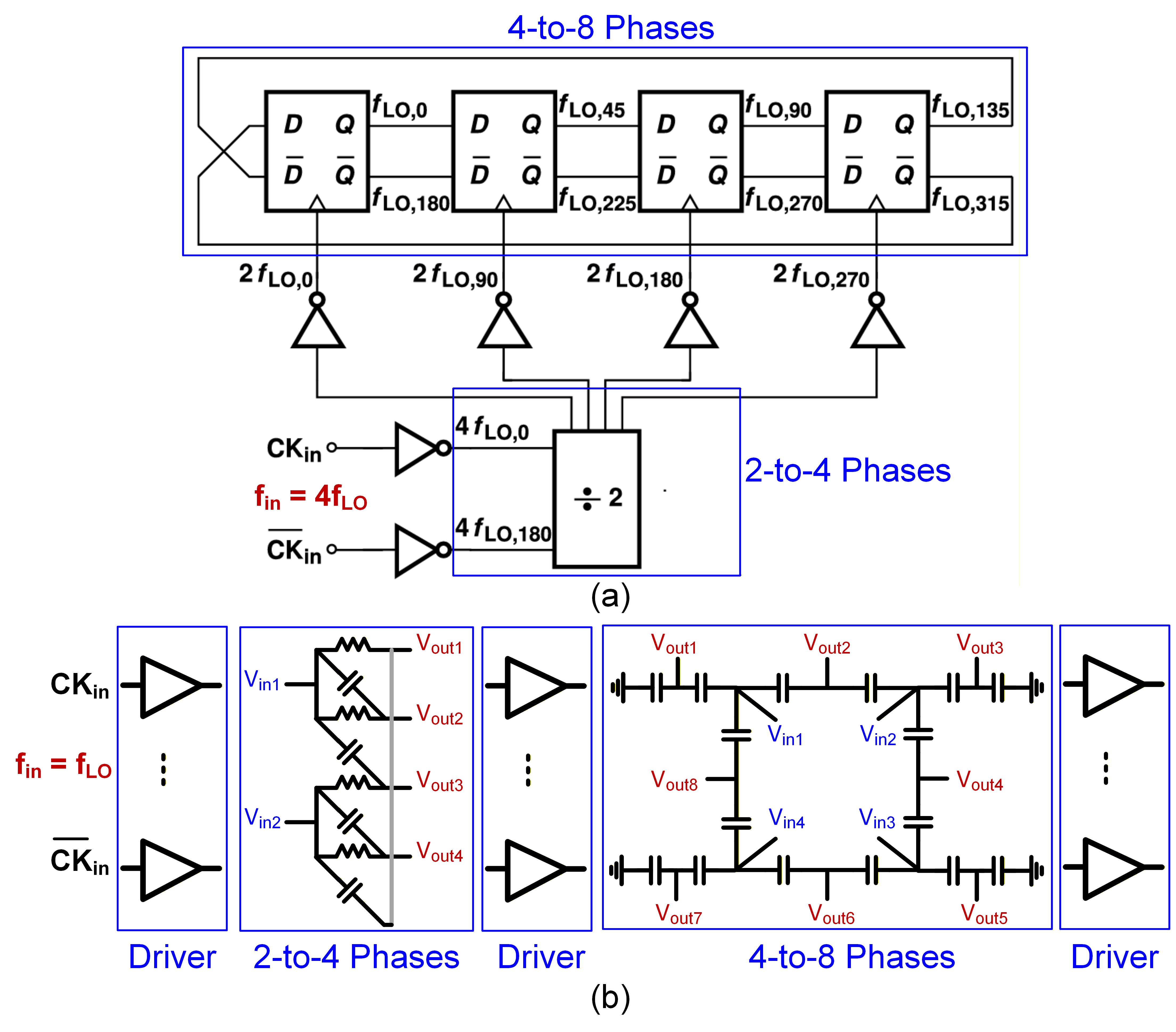}
  \caption{Multiphase clock generation: (a) divider-based generation of eight non-overlapping phases with $f_{in}=4f_{LO}$ and (b) overlapping phase generation using a passive polyphase network, phase interpolation, and RF drivers with $f_{in}=f_{LO}$.}
  \label{fig:CK_Generator}
\end{figure}

A mixer-first RX usually operates using non-overlapping CK phases generated by well-established circuits in lower RF bands. A four-phase CK can be generated using a divide-by-two ($\div 2$) frequency divider comprising two D flip-flops, with an input CK frequency of $f_{in} = 2 f_{CK}$. These D flip-flops can be implemented using static latch circuits at lower frequencies or current-mode logic (CML) circuits at higher frequencies. An eight-phase CK generator \cite{park_jssc_2014, razavi_jssc_2022} is shown in Fig. \ref{fig:CK_Generator}(a), where an input CK at $f_{in} = 4 f_{CK}$ is first applied to a $\div 2$ circuit to generate four CK phases at $2 f_{CK}$. These four phases are then applied to another $\div 2$ circuit comprising four D flip-flops to generate eight CK phases at $f_{CK}$. 

This CK generation approach can be used in FR3 up to a certain frequency, but at the cost of higher power consumption. A 28-nm CMOS implementation of this architecture, with some additional retiming circuits, consumes 7 mW at 1 GHz (7 mW/GHz) and 28 mW at 6 GHz (4.7 mW/GHz), accounting for 30\% and 57\% of the total RX power consumption, respectively \cite{razavi_jssc_2022}. This implies that CK generation can dominate the RX power consumption at higher frequencies within FR3 bands. The required input CK frequency is 
\begin{equation}
    \label{fin_ck}
    f_{in} = \frac{N}{2} f_{CK},
\end{equation}
which can become extremely high in FR3 (e.g., 40 GHz for an eight-phase 10-GHz CK). The higher CK frequency also increases sensitivity of CK distribution network to noise and, as a result, the CK signal will suffer higher jitter in FR3 bands \cite{Mo_sscs_2021, Razavi_tcas1_2021}. Another key limitation is the achievable rise/fall time of the inverter buffers driving the mixer switches, which also impacts the CK jitter. 

\begin{figure}[!t]
  \centering
  \includegraphics[width = 0.7\columnwidth]{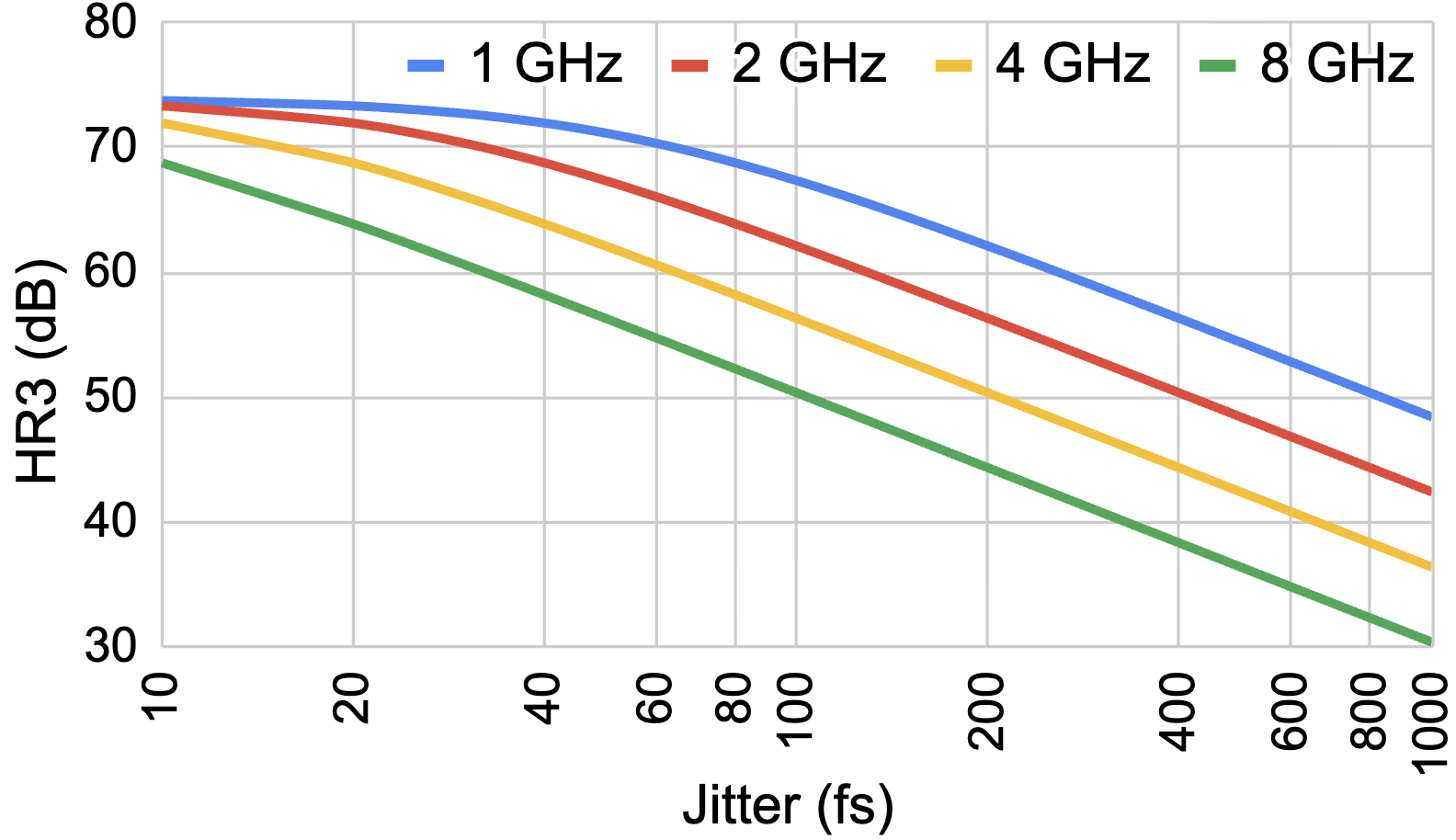}
  \caption{Third-harmonic rejection of the HRM versus the CK jitter.}
  \label{fig:HR3_jitter}
\end{figure}

The impacts of CK phase and amplitude mismatches on harmonic rejection of an HRM have been investigated in \cite{Weldon_jssc_2001, ru_jssc_2009, razavi_sscs_2025}. The third-harmonic rejection for an eight-phase double-balanced HRM can be expressed as
\begin{equation}
    \label{HR3_mismatch}
    {\rm HR_3} =  \frac{\tan^2(\frac{\pi}{8})}{(\frac{\sigma_A}{12})^2 + (\frac{\sigma_{\phi}}{4})^2},
\end{equation}
where $\sigma_A$ and $\sigma_{\phi}$ denote standard deviation of the amplitude and phase mismatch, respectively. The phase mismatch can be related to the jitter as $\sigma_{\phi} = (2\pi f) \sigma_j$. In Fig. \ref{fig:HR3_jitter}, $\rm HR_{3}$ versus CK jitter is shown for frequencies in the range of 1--8 GHz, which their third harmonics can lie in FR1 and FR3 bands. A gain mismatch of $\sigma_A = { 0.1\%}$ has been assumed in (\ref{HR3_mismatch}). It is observed that increasing the frequency results in lower $\rm HR_{3}$ at the same jitter levels. Additionally, at higher frequencies, the roll-off of $\rm HR_{3}$ with jitter starts at lower levels. 

Assuming $\sigma_A \ll \sigma_{\phi}$ in (\ref{HR3_mismatch}), an upper bound on $\rm HR_3$ for a given jitter level can be derived.
In the phase-error-dominated region, $\rm HR_{3}$ decreases by approximately 6~dB when either frequency or rms timing jitter is doubled. Maintaining the same mismatch-limited rejection while doubling frequency therefore requires approximately half the timing jitter under the assumptions of (\ref{HR3_mismatch}). A model developed in \cite{Razavi_tcas1_2021} relates voltage-controlled-oscillator (VCO) power to jitter as
\begin{equation}
    \label{P_VCO_jitter}
    P_{\rm VCO} \propto \frac{1}{\sigma_j^{4}}.
\end{equation}
Therefore, the VCO power consumption should be increased by a factor of 16 to halve the CK jitter. This is not a universal PLL scaling law, but it illustrates why a harmonic-rejection target can impose a substantial CK
power cost at higher frequency. This discussion can be concluded by the insight that power consumption of the mixer-first RXs in FR3 bands can be dominantly determined by that of the CK generation circuits.

The jitter levels considered here refer to the final CK signal applied to the switches. The input reference CK operates at a higher frequency $f_{ {in}} = 4f_{ {LO}}$, and passes through the multi-phase CK generation and then CK distribution circuits which can impose more stringent jitter requirements on the reference CK. 
PLLs with rms jitter below 60~fs at frequencies from 7 to 31~GHz have been reported, including 20.9~fs at 20~GHz in 28-nm CMOS \cite{Zhao_jssc_2023}. The final switch waveform also includes jitter and phase error added by the multiphase generator, buffers, and distribution network; reference-PLL jitter alone is therefore insufficient to predict RX performance.

\subsection{Overlapping CK Generation}

Multi-phase CK signals can also be generated using passive circuits, such as polyphase filters, quadrature couplers, and phase interpolators \cite{Ahmed_tmtt_2024}, followed by active buffers and drivers \cite{razavi_sscs_2025}. This approach, illustrated in Fig. \ref{fig:CK_Generator}(b), can generate overlapping phases directly at the LO frequency, $f_{in}=f_{LO}$, and avoids the higher-frequency input required by a divider-based eight-phase generator. This can significantly reduce power consumption of the LO generation circuits. 

In a mixer-first RX, overlap between CK phases causes loss of charge stored on the capacitors, degrading gain and NF of the RX \cite{Andrews_ESSCIRC_2012, Huang_rfic_2021, Huang_cicc_2024, Iotti_jssc_2020, Anton_jssc_2026}. An LO overlap suppression technique presented in \cite{Huang_rfic_2021} uses inductors between the RF input and two sets of mixers to create a high-impedance path between overlapping switches. This RX uses eight CK phases with a 25\% duty cycle (twice that of the non-overlapping case) and 12.5\% overlap between consecutive phases. Implemented in 45-nm SOI, the RX achieves an NF of 2.4--4.7 dB across 3.7--6.5 GHz, which is competitive with RXs using non-overlapping CK phases \cite{Huang_rfic_2021}.

Another approach, presented in \cite{Anton_jssc_2026}, uses an on-chip harmonically enhanced inverse Class-F quadrature oscillator to generate CK phases with reduced duty cycle and overlap for driving the mixer switches. In addition, a reactive network is introduced between the I and Q mixers to suppress overlap losses. Implemented in 16-nm FinFET, this mixer-first RX achieves an NF of 4.2 dB across 12.3–14.5 GHz FR3 band.


\section{Insights for FR3 Receiver Design}
\label{section:FR3_insights}

The higher frequencies and wider channel bandwidths of FR3 introduce new challenges, as well as opportunities for innovation, in RX circuit design. In this section, we elaborate on the key insights developed throughout the preceding sections.

The frequency translational feature of the mixer-first RX can be leveraged to realize high-performance BB circuits without the complexity, losses, and parasitic limitations associated with RF circuits in FR3. Selectivity can be enhanced using higher-order BB impedance, linearity can be improved through BB amplifier linearization, and NF can be reduced through BB noise cancellation. This insight can be used as a principle to develop new BB circuits for mixer-first RX in FR3 bands. 

The impact of parasitics on RX performance is more significant in FR3. Input parasitic capacitance introduces frequency-dependent losses at the RX input that increase in FR3, leading to higher NF. In addition, the on-state resistance of mixer switches cannot be reduced arbitrarily at high frequencies, as it increases loss through switch parasitic capacitance and raises power consumption in the LO drivers. These constraints limit the achievable NF and linearity in FR3.

Electromagnetic structures such as inductors and transformers can offer several advantages in FR3 compared to FR1, including higher Q factor, lower parasitic capacitance, higher self-resonance frequency, and smaller chip area. These benefits can be leveraged to develop new RX architectures for parasitic resonance, impedance matching, harmonic rejection, and CK overlap compensation. A promising solution for FR3 is to use an input transformer at the RX front-end to absorb switches parasitic capacitance, provide passive voltage gain, and enable the use of smaller switches through impedance transformation.

Harmonic rejection in FR3 should be realized using a minimal number of switches and CK phases to reduce parasitic capacitance and losses. In addition, electromagnetic structures can be used to develop new circuits that function as harmonic traps. The harmonic rejection requirements in FR3 are also expected to be less stringent than in FR1, since most harmonics fall outside the FR3 spectrum.

Mixer-first RXs using overlapping CK phases can achieve higher performance in FR3, where conventional non-overlapping CK generation circuits can become ineffective due to speed limitations or require prohibitively high power consumption. The effects of CK overlap must be compensated through reactive input networks and CK waveform-shaping techniques. Electromagnetic structures can be beneficial in both approaches to realize more effective circuits.

The mixer-first RX benefits from CMOS technology scaling primarily through the availability of high-speed mixer switches with lower parasitic capacitance, as well as CK generation circuits with higher operating frequencies and lower power consumption. However, the reduced supply voltage in advanced CMOS technologies can be insufficient for CK circuits to properly drive the mixer switches. This issue can be addressed using CK boosting techniques, at the cost of extra power consumption.


\section{Conclusion}
\label{section:conclusion}

The design of blocker-tolerant mixer-first RXs for 6G communications in FR3 bands presents a new landscape of both challenges and opportunities. The main challenges arise from the higher frequencies and wider modulation bandwidths of FR3 signals, which can degrade RX circuit performance, as well as the fragmented nature of FR3 bands, which requires both RX and CK generation circuits to be tunable across a broad frequency range. In this paper, the main circuit techniques developed to enhance mixer-first RX performance were explored, their limitations in FR3 were evaluated, and potential solutions at both the circuit and architecture levels were proposed. Opportunities for developing new RX architectures using electromagnetic structures were also discussed to overcome the limitations of existing RX architectures. Finally, design insights for FR3 blocker-tolerant RX were presented.

\bibliographystyle{IEEEtran}
\bibliography{ref_revised}

\end{document}